\documentclass[10pt,twocolumn]{IEEEtran} %to change to the final version with two columns

\usepackage{easyReview}
\usepackage{theorem} \usepackage{cite} \usepackage{stfloats} \usepackage{epsfig} \usepackage{verbatim}
\usepackage{graphicx} \usepackage{amssymb} \usepackage{amsmath} \usepackage{color}
\usepackage{bm}
\usepackage{xcolor}
\usepackage{algorithm}
\usepackage{algorithmicx}

\usepackage{multirow} % 当表格里面的某一项占据多行或者多列时需要导入的包
\usepackage{siunitx}
\usepackage{booktabs}
\usepackage{graphicx} %用于自适应调整表格长宽的包
\usepackage{booktabs} %三线表所用包
\usepackage{array}
\usepackage{makecell}
\usepackage{cellspace}
\usepackage{tabularx}
\newcolumntype{L}[1]{>{\raggedright\arraybackslash}p{#1}}
\newcolumntype{C}[1]{>{\centering\arraybackslash}p{#1}}
\newcolumntype{R}[1]{>{\raggedleft\arraybackslash}p{#1}}

\usepackage{graphicx}
\usepackage{subfigure}
\usepackage{float}
\usepackage{placeins}
\usepackage{subfig}
\usepackage{overpic}
\usepackage{epstopdf}
\usepackage{CJKutf8}
\usepackage{indentfirst}
\usepackage{amsmath}
\usepackage{enumitem}
\usepackage{textcomp}
\usepackage{gensymb}
\usepackage{makecell}
\usepackage{subfigure}
\usepackage{subcaption}

\usepackage{ifthen}

\usepackage{titlesec}

\newboolean{insertCitation}
\setboolean{insertCitation}{true} % 设置为true时插入引用，false时不插入

\newcommand{\conditionalCite}[1]{%
    \ifthenelse{\boolean{insertCitation}}{\cite{#1}}{}%
}

\usepackage{easyReview}
\usepackage{theorem} \usepackage{cite} \usepackage{stfloats} \usepackage{epsfig} \usepackage{verbatim}
\usepackage{graphicx} \usepackage{amssymb} \usepackage{amsmath} \usepackage{color}
\usepackage{bm}
\usepackage{xcolor}
\usepackage{algorithm}
\usepackage{algorithmicx}

\usepackage{multirow} % 当表格里面的某一项占据多行或者多列时需要导入的包
\usepackage{siunitx}
\usepackage{booktabs}
\usepackage{graphicx} %用于自适应调整表格长宽的包
\usepackage{booktabs} %三线表所用包
\usepackage{array}
\usepackage{makecell}
\usepackage{tabularx}

\usepackage{graphicx}
\usepackage{subfigure}
\usepackage{float}
\usepackage{placeins}
\usepackage{subfig}
\usepackage{overpic}
\usepackage{epstopdf}
\usepackage{CJKutf8}
\usepackage{indentfirst}
\usepackage{amsmath}
\usepackage{enumitem}
\usepackage{textcomp}
\usepackage{gensymb}
\usepackage{makecell}
\usepackage{subfigure}
\usepackage{subcaption}

\usepackage{ifthen}

\usepackage{titlesec}

\usepackage{array} % 用于更复杂的列格式化
\usepackage{makecell} % 用于多行单元格内容

\usepackage{wrapfig} % 加载 wrapfig 包

\begin{document}

% Put a box at the end of the proof
\def\QEDclosed{\mbox{\rule[0pt]{1.3ex}{1.3ex}}}
\def\QEDopen{{\setlength{\fboxsep}{0pt}\setlength{\fboxrule}{0.2pt}\fbox{\rule[0pt]{0pt}{1.3ex}\rule[0pt]{1.3ex}{0pt}}}}
\def\QED{\QEDopen}
\def\proof{}
\def\endproof{\hspace*{\fill}~\QED\par\endtrivlist\unskip}

\title{Port-LLM: A Port Prediction Method for Fluid Antenna based on Large Language Models}
\author{Yali~Zhang,
    Haifan~Yin,~\IEEEmembership{Senior Member,~IEEE},
    Weidong~Li,
    Emil~Bj\"{o}rnson,~\IEEEmembership{Fellow,~IEEE}
    and M\'{e}rouane~Debbah,~\IEEEmembership{Fellow,~IEEE}
\thanks{Yali~Zhang, Haifan~Yin and Weidong~Li are with School of Electronic Information and Communications, Huazhong University of Science and Technology, 430074 Wuhan, China (email: yalizhang@hust.edu.cn; yin@hust.edu.cn; weidongli@hust.edu.cn).
}%
\thanks{E. Bj\"ornson is with the Division of Communication Systems,  KTH Royal Institute of Technology, Stockholm, Sweden. E-mail: emilbjo@kth.se.}%
\thanks{M. Debbah is with KU 6G Research Center, Department of Computer and Information Engineering, Khalifa University, Abu Dhabi 127788, UAE (email: merouane.debbah@ku.ac.ae) and also with CentraleSupelec, University Paris-Saclay, 91192 Gif-sur-Yvette, France.}%
\thanks{The corresponding author is Haifan~Yin.}%
\thanks{This work was supported by the Fundamental Research Funds for the Central Universities and the National Natural Science Foundation of China under Grant 62071191. E. Bj\"{o}rnson was supported by the Grant 2022-04222 from the Swedish Research Council.}%
}

\maketitle

\begin{abstract} 
The objective of this study is to address the mobility challenges faced by user equipment (UE) through the implementation of fluid antenna (FA) on the UE side. This approach aims to maintain the time-varying channel in a relatively stable state by strategically relocating the FA to an appropriate port. To the best of our knowledge, this paper introduces, for the first time, the application of large language models (LLMs) in the prediction of FA ports, presenting a novel model termed Port-LLM. Our proposed method for predicting the moving port of the FA is a two-step prediction method. To enhance the learning efficacy of our proposed Port-LLM model, we integrate low-rank adaptation (LoRA) fine-tuning technology. Additionally, to further exploit the natural language processing capabilities of pre-trained LLMs, we propose a framework named Prompt-Port-LLM, which is constructed upon the Port-LLM architecture and incorporates prompt fine-tuning techniques along with a specialized prompt encoder module. The simulation results show that our proposed models all exhibit strong generalization ability and robustness under different numbers of base station antennas and medium-to-high mobility speeds of UE. In comparison to existing methods, the performance of the port predicted by our models demonstrates superior efficacy. Moreover, both of our proposed models achieve millimeter-level inference speed.\par

\end{abstract}

\begin{IEEEkeywords}
Fluid antennas, large language models, channel prediction, moving port prediction, Port-LLM, Prompt-Port-LLM.
\end{IEEEkeywords}

%%%%%%%%%%%%%%%%%%%%%%%%%%%%%%%%%%%%%%%

%%%%%%%%%%%%%%%%%%%%%%%%%%%%%%%%%%%%%%% %\IEEEPARstart{S}{ince} %\newpage

\section{Introduction}\label{sec_intro}
In recent decades, the implementation of multiple-input multiple-output (MIMO) technology has significantly enhanced the capacity and reliability of communication systems. Nevertheless, the fixed deployment of current antennas limits the utilization of spatial degrees of freedom (DoF), particularly in environments with little scattering. Additionally, the constraints imposed by the antenna spacing, i.e., the half-wavelength limitation, further restrict the number of antennas that can be deployed within a confined space. This limitation is particularly pronounced on the user equipment (UE) side, where the spatial dimensions are typically small, resulting in inadequate exploitation of narrow space diversity \cite{FAS_2021,port_selection_2022,performance_limit_2020}. In contrast, a novel antenna technology known as the fluid antenna (FA) or movable antenna, depending on the hardware implementation, offers the capability to switch among various positions, referred to as ``ports", within a specified area. 
%In this paper, the terms ``fluid antenna" and ``movable antenna" are considered interchangeable. In the subsequent discussion, we will employ the term ``fluid antenna" to encompass both concepts. 
The position and configuration of the FA can be dynamically adjusted, allowing for greater adaptability \cite{lq_antenna_2021, zhu2024historical, 6D_movable_antenna_1, 6D_movable_antenna_2}. Despite the limited size of the movable area of the FA, the potential for numerous movable ports allows for significant diversity gain across a multitude of spatially dependent ports. Furthermore, the continuous mobility of the FA within a specified area facilitates the optimal exploitation of spatial DoF, thereby enhancing the conditions of the wireless channel.\par

%%FA is implemented by using flexible antenna architectures such as liquid antennas (including liquid metal, water-based and non-water-based materials), reconfigurable radio frequency (RF) pixel-based antennas, and stepper motor-based antennas to reconfigurable antenna positions so that the radiating antenna can be located anywhere in a given area. 

The fluid antenna system (FAS) presents significant advantages and potential applications \cite{zhu2023movable}. Notably, the integration of FAS with MIMO technology, referred to as MIMO-FAS, has the capacity to enhance the performance of a MIMO system by selecting ports that enhance the beamforming gain or improve the MIMO rank conditions, thereby facilitating exceptionally high data transmission rates and enhanced reliability. In multi-user scenarios, FAS can be employed for interference suppression purposes, allowing users to leverage naturally occurring interference nulls in the propagation environment by adjusting the FA port, which reduces the need for interference-suppression precoding at the BS. Furthermore, FAS can be integrated with reconfigurable intelligent surface (RIS) technology, thereby circumventing the need for intricate optimization processes associated with RIS \cite{FAS_1, FAS_2}.\par

FAS also encounters numerous challenges \cite{zhu2023movable}, one of which pertains to the selection of ports. The advantages of FAS compared to conventional fixed-location antenna systems (FPAs) are primarily due to its flexible antenna positioning capabilities. Nevertheless, identifying the appropriate port for the FA to achieve superior communication performance presents a significant challenge, 
%as the channel response is characterized by a highly nonlinear relationship with the position of the FA. 
as the channel response exhibits a significant degree of nonlinearity as a function of the spatial positioning of the FA. Conventional optimization techniques for FAS port selection encompass the gradient descent \cite{antenna_position_op_2024} method, successive convex approximation (SCA) \cite{mimo_capacity_2024}, alternating optimization (AO) method, exhaustive search method, etc. Nevertheless, these approaches either necessitate precisely known channel state information (CSI) or entail significant time and computational resources to identify an appropriate port.\par
%Nonetheless, these approaches either necessitate complete knowledge of CSI or incur significant time and computational costs in the pursuit of the optimal solution.\par

The issue of mobility, referred to as the curse of mobility \cite{yin2020addressing}, has consistently been a significant concern within the field of communications. The movement of the UE can introduce a significant Doppler effect, leading to the obsolescence of the communication channel. If the BS performs precoding design with the outdated CSI, it may result in a decline in system performance. To address this mobility challenge, the work in \cite{yin2020addressing} proposed the Vec Prony method and the Prony
based angular-delay domain (PAD) method for channel prediction. Their findings indicate that it can achieve performance levels comparable to stationary scenarios with unaltered channels. As research in FAs has progressed, the paper \cite{Li2024TransformingTT} proposed utilizing the FA to mitigate mobility issues. Their study introduces a matrix pencil-based moving port (MPMP) prediction method, which facilitates the port selection for FA. 
%By dynamically adjusting the position of the FA on the UE side to the optimal port, the time-varying channel approximation can be maintained at a constant level. 
Empirical results demonstrate that the MPMP method outperforms the Vec Prony algorithm in both medium and high mobility scenarios. %However, the MPMP algorithm is influenced by the port resolution and the quantity of antennas at the BS.\par
%However, the accuracy of the MPMP algorithm is significantly influenced by the port resolution within the movable area of FA. An increase in the number of ports leads to heightened time and computational complexity. Furthermore, as the number of antennas on the BS side increases, the time and computational demands associated with employing the MPMP algorithm for mobile port selection on the UE side also escalate considerably.\par

Deep learning (DL) technology has garnered significant interest within the domain of wireless communication, owing to its robust capabilities in feature extraction and modeling \cite{LLM_Telecom_2024}. This has led to notable advancements, including the application of DL for tasks such as channel prediction, solving beamforming vector \cite{superdirective_beamforming_2024}, antenna selection (AS) \cite{machine_port_selection_2016}, etc. From these applications, it is evident that applying DL technology to FA port prediction to address mobility issues has great potential, since wireless channels have much structure even if it is hard to model. According to the research conducted by the paper \cite{Li2024TransformingTT}, the port selection for FA is inherently a time-sensitive issue. Techniques such as recurrent neural networks (RNNs) \cite{Lipton2015ACR}, long short-term memory (LSTM) \cite{graves2012long} network, and gated recurrent unit (GRU) \cite{Chung2014EmpiricalEO} are frequently employed to tackle problems characterized by temporal variations. However, these traditional neural network models designed for time series analysis typically exhibit a small architecture and possess a limited number of internal learnable parameters. Consequently, their capacity to model complex problems is constrained. Furthermore, these models demonstrate limited generalization capabilities and exhibit heightened sensitivity to variations in environmental parameters, thereby constraining their practical applicability in real-world scenarios.\par

Large language models (LLMs), such as GPT-1, GPT-2 \cite{radford2019language}, GPT-3, and LLaMa, have significantly transformed the fields of natural language processing (NLP) and artificial intelligence (AI). 
%%The robust modeling and generalization capabilities inherent in LLMs present substantial opportunities for their application within the telecommunications sector []. 
Compared to other neural network-based models, models based on LLMs not only possess excellent NLP capabilities but also have inherent sequence modeling capabilities acquired during the pre-training phase.
Currently, researchers have initiated investigations into the application of LLMs within the physical layer of wireless communication networks. Notable examples include the Csi-LLM \cite{Fan2024CsiLLMAN} and LLM4CP \cite{LLM4CP_2024} models, which have been developed for channel prediction, as well as proposals for utilizing LLMs in beam prediction tasks \cite{beam_prediction_LLM_2024}. However, there is a notable gap in the literature regarding the application of LLMs for port prediction in FA. This paper addresses this gap by proposing an FA port prediction model that leverages LLMs. However, given that the extensive datasets utilized for the pre-training of LLMs predominantly consist of diverse textual data, these models lack the capability to interpret wireless communication data. As a result, aligning wireless communication data with natural language data remains a significant challenge when implementing LLMs within the physical layer of wireless communication networks.\par

In this article, we introduce an FA port prediction model, designated as Port-LLM, to tackle the challenges associated with mobility. Our objective is to maintain a relatively stable channel despite the movement of the UE by relocating the FA to the port predicted by our model at each moment. The foundational architecture of our model is derived from the pre-trained GPT-2 \cite{radford2019language} framework, and we employ low-
rank adaptation (LoRA) \cite{Hu2021LoRALA} technology to fine-tune the loaded GPT-2 model. The research presented in \cite{Li2024TransformingTT} indicates that the critical factor in selecting ports for FA lies in the precise forecasting of CSI pertaining to all movable ports of the FA on the UE side at any given moment. As a result, our proposed LLM-based FA port prediction model utilizes the CSI corresponding to all movable ports on the UE side from the preceding $T$ moments, referred to as the ``channel tables," as input. Subsequently, the proposed Port-LLM is employed to forecast the channel tables for the subsequent $F$ moments. Ultimately, the moving ports of the FA at the subsequent $F$ moments are derived from the channel tables predicted by our model, in conjunction with the known reference channels that require alignment. Generally, the process of utilizing the proposed Port-LLM for FA port prediction can be categorized into two primary phases: the first phase involves channel prediction, while the second phase pertains to port selection.\par

Furthermore, we propose a model termed Prompt-Port-LLM, which is based on prompt fine-tuning. This model distinguishes itself from our previously proposed Port-LLM by employing dynamic prompts for the fine-tuning of the LLMs, whereas the Port-LLM utilizes LoRA fine-tuning for the same purpose. Based on retaining the modeling capabilities of the pre-trained LLMs used by the Port-LLM model, our proposed Prompt-Port-LLM model further explores its potential in NLP. In the proposed Prompt-Port-LLM model, we design specialized dynamic prompts and a specialized prompt encoding module to assist in training. The dynamic prompts can provide real-time task knowledge based on the input data, while the prompt encoding module transforms static prompts into optimizable continuous embeddings, thereby further enhancing the performance of our model. The main contributions of this paper are as follows:
\begin{itemize}
    \item This paper represents the first application of LLMs to the task of FA port selection, introducing an innovative LLM-based FA port prediction model, referred to as Port-LLM. Our model utilizes the channel tables associated with all movable ports of the FA over a preceding period of $T$ time intervals as input, and subsequently predicts the moving ports of the FA for the forthcoming $F$ time intervals. By repositioning the FA to the port predicted by our model at each time, we aim to keep the time-varying channel approximately constant.\par
    
    \item To ensure that the wireless communication data pertinent to this task is aligned with the data patterns of the pre-trained LLM, we develop the specialized data processing module, input embedding module and output projection module. Meanwhile, we conduct LoRA fine-tuning on the GPT-2 model. By employing low-rank matrix techniques, the LoRA fine-tuning approach significantly reduces the number of parameters required for retraining our model on this specific task compared to the full fine-tuning technique, while preserving the knowledge acquired during the pre-training phase of the GPT-2 model.\par

    \item In order to further leverage the NLP capabilities of pre-trained LLMs, we also propose a framework termed Prompt-Port-LLM, which is built upon the Port-LLM architecture and incorporates prompt fine-tuning technique. Within the Prompt-Port-LLM framework, we design the specialized dynamic prompts and a prompt encoding module to assist in model training, thereby further enhancing model performance.\par

\end{itemize}

\emph{Notation:} We use boldface to denote matrices and vectors. $\mathbb{R}^n$ and $\mathbb{C}^n$ denote the spaces of $n$-dimension real and complex numbers, respectively. $\left( \cdot \right) ^T$ represents the transpose. $\text{argmin} \left( \cdot \right) $ refers to the input parameter that minimizes the objective function. $\text{unravel\_index}\left( p,\left( n,m \right) \right) $ denotes the multi-dimensional coordinates associated with the integer $p$ within an $n\times m$ dimensional matrix. $\left| \cdot \right|$ is the absolute value, $\lVert \cdot \rVert $ represents the Euclidean norm, and $\mathbb{E}\left[ \cdot \right] $ represents the expectation operator. $\lceil \cdot \rceil$ is the rounding up operation.\par

\section{System model}\label{sec_system_model}
\begin{figure}[h]
    \centering
    \includegraphics[scale=0.55]{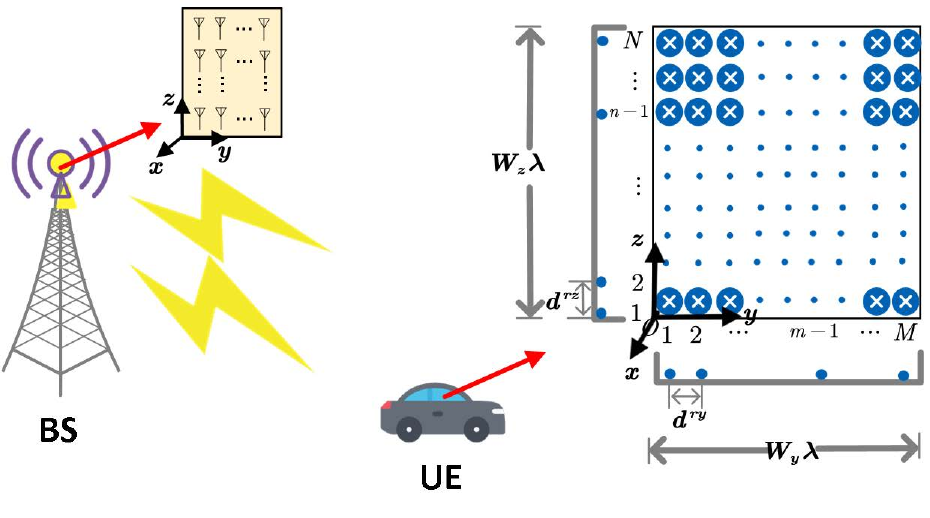}
    \caption{The FA-assisted DL wireless communication system.}
    \label{fig:BE_UE}
\end{figure}

We consider a time division duplexing (TDD) system, where a BS with an $N_{y}\times N_{z}$ uniform planar array (UPA) serves a certain UE that is equipped with an FA. The UE-side FA can be dynamically reconfigured by mechanical actuation or electronic switching control, allowing its radiating unit to quickly switch operating port positions within its preset movable area. Fig. \ref{fig:BE_UE} is the downlink (DL) wireless communication system with FA at the UE side.\par
The placement of the antennas on the BS side is static, with the antenna panel situated within the $yOz$ plane. The spacing between the antennas on the panel, along the $y$-axis and $z$-axis, is denoted by $d^{ty}$ and $d^{tz}$, respectively. Conversely, the UE-side antenna is capable of movement within the $yOz$ plane, characterized by a moving area of $W_y\lambda \times W_z\lambda$, where $W_y\lambda$ and $W_z\lambda$ denote the permissible displacements along the $y$-axis and $z$-axis, respectively. The symbol $\lambda =\frac{c}{f_c}$ denotes the wavelength, while $c$ and $f_c$ represent the speed of light and the carrier frequency, respectively. It is assumed that, on the UE side, the quantities of movable antenna ports along the $y$-axis and $z$-axis are denoted by $M$ and $N$, respectively. The inter-port distances are defined as follows:
\begin{equation}
    \label{eq:inter-port-y}
    d^{ry}=\frac{W_y\lambda}{M-1}=\frac{\lambda}{\rho _y},
\end{equation}
\begin{equation}
    \label{eq:inter-port-z}
    d^{rz}=\frac{W_z\lambda}{N-1}=\frac{\lambda}{\rho _z},
\end{equation}
where $\rho _y=\frac{M-1}{W_y}$ and $\rho _z=\frac{N-1}{W_z}$ are utilized to represent the port density along the $y$-axis and $z$-axis, respectively.\par

On the BS side, the coordinate position vector of the $k$-th antenna is
\begin{equation}
    \label{eq:1}
     \mathbf{d}^{\textrm{tx}}_{k}=\left[ 0,d^{ty}\left( n_y-1 \right) ,d^{tz}\left( n_z-1 \right) \right] ^T,
\end{equation}
where $1\le n_y\le N_y,1\le n_z\le N_z$. And $1\le k\le N_t, N_t=N_y\times N_z$.\par
On the UE side, the coordinate position vector of the antenna located at the $\left(n,m\right)$-th port is represented as
\begin{equation}
    \label{eq:2}
    \mathbf{d}_{n,m}^{\textrm{rx}}=\left[ 0,d^{ry}\left( m-1 \right) ,d^{rz}\left( n-1 \right) \right] ^T,
\end{equation}
where $1\le m\le M, 1\le n\le N$. The spherical unit vectors on the BS and UE sides are:
\begin{equation}
    \label{eq:3}
    \mathbf{r}^{\textrm{tx}}=\left[ \begin{array}{c}
	\sin \theta _{\text{EOD}}\cos \phi _{\text{AOD}}\\
	\sin \theta _{\text{EOD}}\sin \phi _{\text{AOD}}\\
	\cos \theta _{\text{EOD}}\\
\end{array} \right] ,
\end{equation}
\begin{equation}
    \label{eq:4}
    \ \mathbf{r}^{\textrm{rx}}=\left[ \begin{array}{c}
	\sin \theta _{\text{EOA}}\cos \phi _{\text{AOA}}\\
	\sin \theta _{\text{EOA}}\sin \phi _{\text{AOA}}\\
	\cos \theta _{\text{EOA}}\\
\end{array} \right] ,
\end{equation}
where $\theta _{\text{EOA}}, \phi _{\text{EOA}}, \theta _{\text{EOD}}, \phi _{\text{EOD}}$ correspond to the elevation angle of arrival (EOA), azimuth angle of arrival (AOA), elevation angle of departure (EOD), and azimuth angle of departure (AOD), respectively. Furthermore, $\theta _{\text{EOA}}, \theta _{\text{EOD}}\in \left[ 0,\pi \right]$ and $ \phi _{\text{AOA}}, \phi _{\text{AOD}}\in \left( -\pi ,\pi \right] $. The Doppler frequency shift is denoted by $w=\frac{\left( \mathbf{r}^{\textrm{rx}} \right) ^T\mathbf{v}}{\lambda}$, where $\mathbf{v}$ is a vector that represents the velocity of the UE.\par

%As indicated in the study \cite{3GPP2019Study}, 
Similar to the model used in the study \cite{3GPP2019Study}, we consider a scenario that encompasses one line-of-sight (LoS) path and $P$ non-line-of-sight (NLoS) paths. Therefore, the channel coefficient between the $k$-th antenna on the BS side and the UE side antenna at the $\left(n,m\right)$-th port at time $t$ can be expressed as:
\begin{equation}
    \label{eq:5}
    \begin{split}
        h_{(k,n,m)}(t) &= \sum_{p=1}^{P+1} \alpha_p \beta_p e^{\frac{j2\pi (\mathbf{r}_p^{\textrm{rx}})^T \mathbf{d}_{n,m}^{\textrm{rx}}}{\lambda}} \\
        &\quad \times e^{\frac{j2\pi (\mathbf{r}_p^{\textrm{tx}})^T \mathbf{d}_k^{\textrm{tx}}}{\lambda}} e^{j2\pi w_p t} e^{j2\pi f \tau_p},
    \end{split}
\end{equation}
where $f$ is the frequency, while $\tau_p$ and $w_p$ denote the delay and Doppler frequency shift of the $p$-th path, respectively. Moreover, $\beta_p$ denotes the amplitude of the $p$-th path and
\begin{equation}
    \label{eq:6}
    \alpha _p=\begin{cases}
	\sqrt{\frac{1}{K_R+1}},&		\ p\in \text{NLoS},\\
	\sqrt{\frac{K_R}{K_R+1}},&		\ p\in \text{LoS},\\
\end{cases}
\end{equation}
where $K_R$ is the Ricean $K$-factor.\par
Furthermore, at time $t$, the channel coefficient between all BS-side antennas and the UE-side antenna located at the $\left(n,m\right)$-th port can be represented as
\begin{equation}
    \label{eq:7}
    \begin{split}
        \mathbf{h}_{(n,m)}(t) &= \left[ h_{(1,n,m)}(t), \cdots, h_{(N_t,n,m)}(t) \right]^T \\
        &= \mathbf{A} \mathbf{c}_{(n,m)}(t) \in \mathbb{C}^{N_t\times 1},
    \end{split}
\end{equation}
where $
\mathbf{A}=\left[ \mathbf{a}\left( \theta _{1}^{\textrm{tx}},\phi _{1}^{\textrm{tx}} \right) ,\mathbf{a}\left( \theta _{2}^{\textrm{tx}},\phi _{2}^{\textrm{tx}} \right) ,\cdots ,\mathbf{a}\left( \theta _{P+1}^{\textrm{tx}},a_{P+1}^{\textrm{tx}} \right) \right] \in \mathbb{C}^{N_t\times \left( P+1 \right)}
$ represents the steering vectors of all paths. $\theta_{p}^{\textrm{tx}}$ and $\phi_{p}^{\textrm{tx}}$ denote the EOD and AOD of the $p$-th path, respectively. The 3-D steering vector of the $p$-th path is defined as
\begin{equation}
    \label{eq:steering}
    \mathbf{a}\left( \theta _{p}^{\textrm{tx}},\phi _{p}^{\textrm{tx}} \right) =\mathbf{a}_y\left( \theta _{p}^{\textrm{tx}},\phi _{p}^{\textrm{tx}} \right) \otimes \mathbf{a}_z\left( \theta _{p}^{\textrm{tx}} \right) \in \mathbb{C}^{N_t\times 1},
\end{equation}
where
\begin{equation}
    \label{eq:steering_y}
    \mathbf{a}_y\left( \theta _{p}^{\textrm{tx}},\phi _{p}^{\textrm{tx}} \right) =\left[ 1,\cdots ,e^{j\frac{2\pi}{\lambda}\sin \theta _{p}^{\textrm{tx}}\sin \phi _{p}^{\textrm{tx}}d^{ty}\left( N_y-1 \right)} \right] ^T,
\end{equation}
\begin{equation}
    \label{eq:steering_z}
    \mathbf{a}_z\left( \theta _{p}^{\textrm{tx}} \right) =\left[ 1,\cdots ,e^{j\frac{2\pi}{\lambda}\cos \theta _{p}^{\textrm{tx}}d^{tz}\left( N_z-1 \right)} \right] ^T.
\end{equation}
Moreover, the vector $\mathbf{c}_{\left( n,m \right)}\left( t \right) \in \mathbb{C}^{\left( P+1 \right) \times 1}$ is given by
\begin{equation}
    \label{eq:8}
    \begin{split}
        \mathbf{c}_{(n,m)}(t) = &\left[ c_{\left(1,n,m\right)} e^{j2\pi w_1 t}, \cdots, \right. \\
        &\left. c_{\left(P+1,n,m\right)} e^{j2\pi w_{P+1} t} \right]^T,
    \end{split}
\end{equation}
where $c_{\left( p,n,m \right)}=c_pe^{j\frac{2\pi}{\lambda}\left[ \sin \theta _{p}^{\textrm{rx}}\sin \phi _{p}^{\textrm{rx}}d^{ry}\left( m-1 \right) +\cos \theta _{p}^{\textrm{rx}}d^{rz}\left( n-1 \right) \right]}$ and $c_p=\alpha _p\beta _pe^{j2\pi f\tau _p}$. $\theta_p^{\textrm{rx}}$ and $\phi_p^{\textrm{rx}}$ are the EOA and AOA of the $p$-th path, respectively.\par

At time $t$, when the UE antenna is positioned at the $\left(1,1\right)$-th port, the mathematical representation of the channel between all BS antennas and the UE antenna is
\begin{equation}
    \label{eq:ref_channel}
    \begin{split}
        \mathbf{h}_{(1,1)}(t) &= \left[ h_{(1,1,1)}(t), \cdots, h_{(N_t,1,1)}(t) \right]^T \\
        &= \mathbf{A} \mathbf{c}_{(1,1)}(t).
    \end{split}
\end{equation}\par
Without loss of generality, we designate $\mathbf{h}_{\left( 1,1 \right)}\left( t \right) $ as the reference channel. At time $\left(t+\varDelta t\right)$, the channel transitions to $\mathbf{h}_{\left( 1,1 \right)}\left( t+\varDelta t \right) $ due to the mobility of the UE-side antenna. \par
% To achieve an approximate invariance of the channel, we adjust the antenna to the $\left(n,m\right)$-th port, resulting in the channel becoming $h_{\left( n,m \right)}\left( t+\varDelta t \right) $, which closely resembles the reference channel $h_{\left( 1,1 \right)}\left( t \right) $.\par
At time $\left( t+\varDelta t \right)$, we represent the channel coefficients associated with all ports as follows:
\begin{equation}
    \label{eq:port_table}
    \mathbf{S}\left( t+\varDelta t \right) =\left\{ \mathbf{S}_1\left( t+\varDelta t \right) ,\cdots ,\mathbf{S}_{N_t}\left( t+\varDelta t \right) \right\} \in \mathbb{C}^{N_t\times N\times M},
\end{equation}
where
\begin{equation}
    \label{eq:port_table_i}
    \mathbf{S}_i\left( t+\varDelta t \right) =\left[ \begin{matrix}{}
    	h_{\left( i,1,1 \right)}\left( t \right)&		\cdots&		h_{\left( i,1,M \right)}\left( t \right)\\
    	\vdots&		\vdots&		\vdots\\
    	h_{\left( i,N,1 \right)}\left( t \right)&		\cdots&		h_{\left( i,N,M \right)}\left( t \right)\\
    \end{matrix} \right] \in \mathbb{C}^{N\times M},
\end{equation}
denotes the channel matrix between the $i$-th antenna at the BS and all ports of the FA at the time $\left(t+\varDelta t\right)$.\par
%And $h_{\left( i,n,m \right)}\left( t+\varDelta t \right) , 1\le i\le N_t, 1\le n\le N,1\le m\le M$ denots the CSI between the antenna located at the UE and the $i$-th antenna situated at the BS when the UE side antenna is positioned at the $\left(n,m\right)$-th port at a given time $\left( t+\varDelta t \right)$.\par

The objective of our study is to identify a specific port, denoted by $\left( n_{\text{opt}},m_{\text{opt}} \right) $, at a future time $\left( t+\varDelta t \right)$, from the entire set of available ports. This selection aims to ensure that when the UE side antenna slides to the $\left( n_{\text{opt}},m_{\text{opt}} \right)$-th port, the channel information $\mathbf{h}_{\left( n_{\text{opt}},m_{\text{opt}} \right)}\left( t+\varDelta t \right)$ closely aligns with the reference channel $\mathbf{h}_{\left( 1,1 \right)}\left( t \right)$, thereby keeping the channel approximate constant. We expand the $\mathbf{h}_{\left( 1,1 \right)}\left( t \right) \in \mathbb{C}^{N_{t}\times 1}$ in order to obtain the reference channel matrix corresponding to all ports as outlined below:
\begin{equation}
    \label{eq:ref_channel_table}
    \mathbf{H}_{\text{ref}}\left( t \right) =\left\{ \mathbf{H}_1\left( t \right) ,\cdots ,\mathbf{H}_{N_t}\left( t \right) \right\} \in \mathbb{C}^{N_t\times N\times M},
\end{equation}
where
\begin{equation}
    \label{eq:ref_channel_tabel_i}
    \mathbf{H}_{i}\left( t \right) =\left[ \begin{matrix}{}
	h_{\left( i,1,1 \right)}\left( t \right)&		\cdots&		h_{\left( i,1,1 \right)}\left( t \right)\\
	\vdots&		\cdots&		\vdots\\
	h_{\left( i,1,1 \right)}\left( t \right)&		\cdots&		h_{\left( i,1,1 \right)}\left( t \right)\\
\end{matrix} \right] \in \mathbb{C}^{N\times M}.
\end{equation}\par

To maintain a relatively static channel, the following expression can be utilized to determine the moving port $\left(n_{\text{opt}},m_{\text{opt}}\right)$ of FA at time $\left(t+\varDelta t\right)$:
\begin{equation}
\begin{aligned}
\left( n_{\text{opt}},m_{\text{opt}} \right) = 
\text{unravel\_index}\bigg( 
&\text{argmin} 
\bigg( 
\sum_{i=1}^{N_t} 
\Big| \mathbf{S}_i\left( t+\varDelta t \right) - \\
&\mathbf{H}_i\left( t \right) \Big| \bigg)
, \left( N,M \right) \bigg)
\end{aligned}.
\end{equation}\par
The aforementioned formula indicates that when the reference channel is known, determining the moving port of FA at the subsequent moment for maintaining a relatively stable channel hinges on acquiring the channel matrices that connect all antennas on the BS side with all movable ports of the FA on the UE side at that particular moment. For the sake of clarity in the subsequent sections, we will call the channel matrix between the $i$-th antenna on the BS side and all movable ports of FA at a given time as a ``channel table".\par

\section{Port-LLM}\label{sec_Port-LLM}
In this section, we propose a model for predicting the moving port of FA, referred to as Port-LLM, which is grounded in LLMs technology. The primary objective of our model is to ensure that the channel remains relatively stable by relocating the FA on the UE side to the anticipated port while the UE is in motion. In the subsequent model parameters, in order to simplify the notation, we set the number of antennas on the BS side $N_t$ to 1. Note that the simulation validations are performed in multi-antenna setting. To accomplish the above objective, we initially employ the proposed Port-LLM model to forecast the channel tables $\mathbf{\hat{S}}=\left\{ \mathbf{\hat{S}}_1,\cdots ,\mathbf{\hat{S}}_F \right\} \in \mathbb{C}^{F\times N\times M}$ for the subsequent $F$ moments by utilizing the channel tables $\mathbf{S}=\left\{ \mathbf{S}_1,\cdots ,\mathbf{S}_T \right\} \in \mathbb{C}^{T\times N\times M}$ from the preceding $T$ moments. Subsequently, we proceed to utilize the predicted channel tables $\mathbf{\hat{S}}\in \mathbb{C}^{F\times N\times M}$ and the known reference channel $\mathbf{H}_{\text{ref}} \in \mathbb{C}^{F\times N\times M}$ to obtain the moving ports $\mathbf{P}=\left\{ \mathbf{p}_1,\cdots ,\mathbf{p}_F \right\}\in \mathbb{R}^{F\times 2\times 1}$ for the subsequent $F$ moments. 
%\add{It is important to highlight that although our proposed neural network model is built based on the data parameters in the SISO scenario, the subsequent simulation validation is performed in the MISO scenario.}\par 

\begin{figure*}[t]
    \centering
    \includegraphics[scale=0.50]{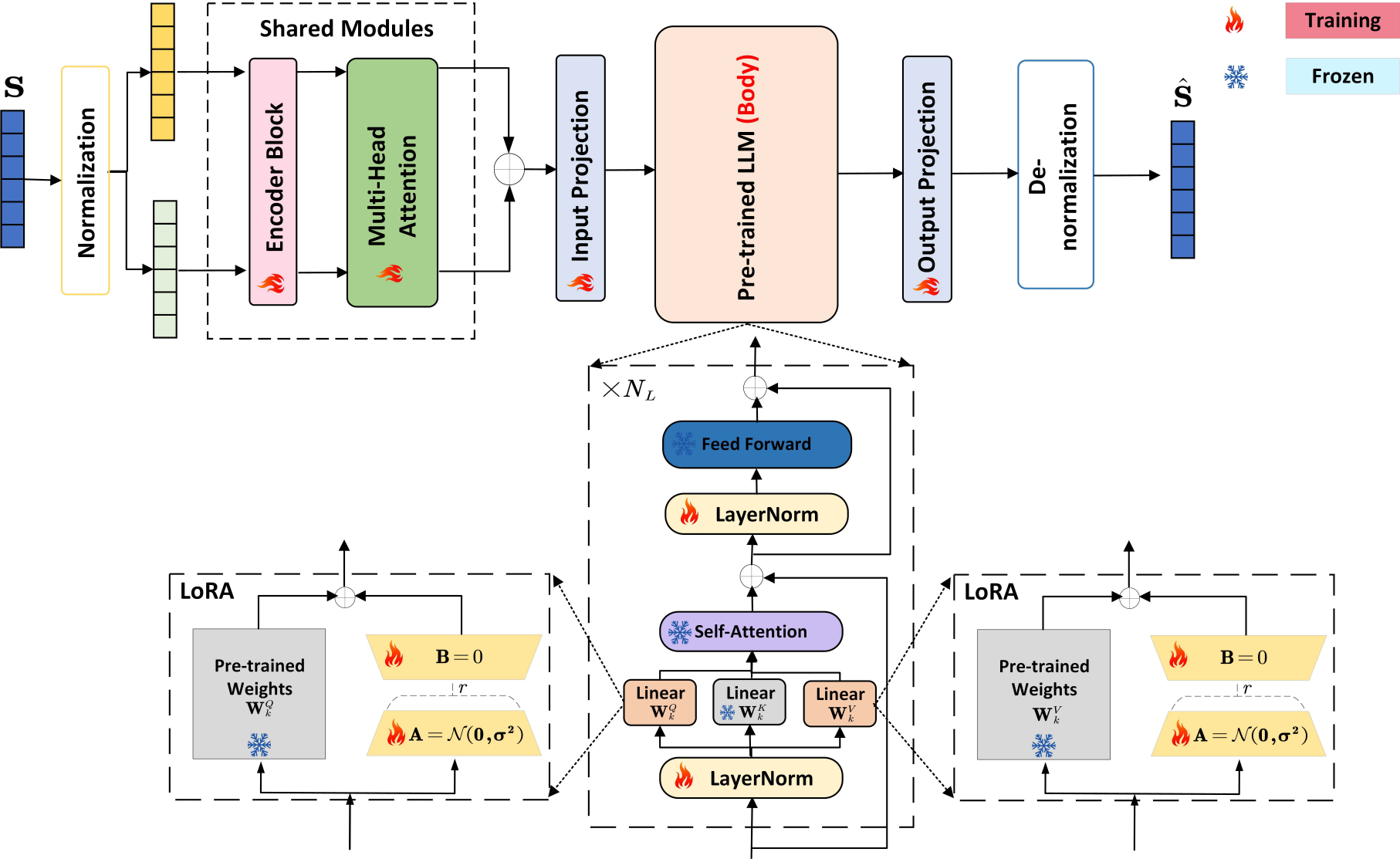}
    \caption{The architecture of our proposed Port-LLM model.}
    \label{fig:PORT-LLM}
\end{figure*}

\begin{figure}
    \centering
    \includegraphics[width=0.95\linewidth]{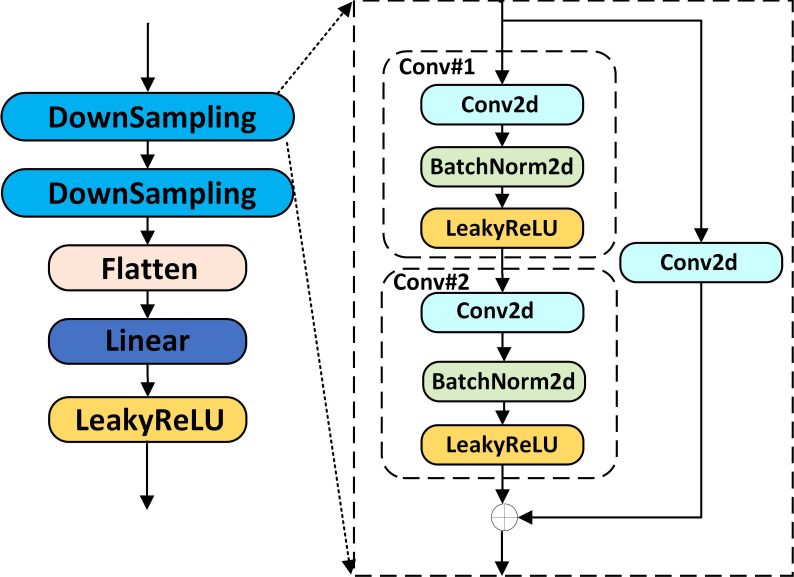}
    \caption{The architecture of the Encoder Block module.}
    \label{fig:enocder-block-1}
\end{figure}

\subsection{Network Architecture}\label{subsec_network}
The network architecture of our proposed Port-LLM model primarily comprises the data processing, input embedding, backbone network, and output projection modules.\par
\subsubsection{Data Preprocessing}\label{data_preprocessing}
To enhance the convergence rate during the training of our model, we initially apply mean-standard deviation normalization to the input data $\mathbf{S}\in \mathbb{C}^{T\times N\times M}$, i.e., $\mathbf{\bar{S}}=\frac{\mathbf{S}-\mathbf{\mu }}{\mathbf{\sigma }}$, where $\mathbf{\mu}$ and $\mathbf{\sigma}$ denote the mean and standard deviation of the input data, respectively. Given that neural network models typically operate on real-valued inputs and the input data $\mathbf{S}$ is complex in nature, we decompose $\mathbf{\bar{S}}\in \mathbb{C}^{T\times N\times M}$ into two components: the real part $\mathbf{\bar{S}}_r\in \mathbb{R}^{T\times N\times M}$ and the imaginary part $\mathbf{\bar{S}}_i\in \mathbb{R}^{T\times N\times M}$.\par

\subsubsection{Input Embedding}\label{input_embedding}
For feature extraction in subsequent modeling, the real part $\mathbf{\bar{S}}_r\in \mathbb{R}^{T\times N\times M}$ and imaginary part $\mathbf{\bar{S}}_i\in \mathbb{R}^{T\times N\times M}$ are first processed separately through a shared feature extraction module (referred to as the ``Shared Module"). This shared module consists of two key components: an encoder module (``Encoder Block") and a multi-head attention module \cite{vaswani2017attention}. The encoder module is responsible for dimensionality reduction and feature compression of the input data, with its detailed architecture illustrated in Fig. 3. The multi-head attention module, on the other hand, is primarily employed to extract spatio-temporal features from both the real and imaginary components. Given that the real and imaginary data undergo identical processing within the ``Shared Module", we focus on the real component as an illustrative example for clarity in the following discussion.\par
The processing sequence for the real part data $\mathbf{\bar{S}}_r\in \mathbb{R}^{T\times N\times M}$ within the Encoder Block module is delineated as follows: initially, the input data is subjected to a sequential passage through two DownSampling modules. Each of these modules reduces the spatial dimensions (length and width) of the data by half, while maintaining a constant number of channels. The detailed procedure for the DownSampling process is illustrated below:\par
\begin{equation}
    \label{eq:downsampling}
    \begin{aligned}
        \mathbf{\mathring{S}}_r
        &=\text{DownSampling}\left( \mathbf{\bar{S}}_r \right) \\
        &=\text{Conv2d}\left( \mathbf{\bar{S}}_r \right) +\text{Conv\#}2\left( \text{Conv\#}1\left( \mathbf{\bar{S}}_r \right) \right),
    \end{aligned}
\end{equation}
where $\mathbf{\mathring{S}}_r\in \mathbb{R}^{T\times \frac{N}{2}\times \frac{M}{2}}$, and the skip connection in the DownSampling block is introduced to mitigate feature loss. On the right-hand side of the equation,\par

\begin{itemize}
    \item The first convolution operation (Conv2d) performs spatial downsampling, halving the length and width of the input data. Its parameters are: kernel size = 1, stride = 2, padding = 1.\par
    \item Both $\text{Conv\#1}$ and $\text{Conv\#2}$ operations consist of a Conv2d, followed by BatchNorm2d and LeakyReLU activation. However, they serve distinct purposes: $\text{Conv\#1}$ operation performs downsampling (halving spatial dimensions) with parameters: kernel size = 3, stride = 2, padding = 1; $\text{Conv\#2}$ operation enhances feature representation while preserving input dimensionality and its parameters are: kernel size = 3, stride = 1, padding = 1.\par
\end{itemize}

Following two DownSampling procedures, the dimensionality of the real data is reduced from an initial value of $T\times N\times M$ to $T\times \lceil \frac{N}{4} \rceil \times \lceil \frac{M}{4} \rceil $. Subsequently, the multidimensional feature data is transformed into a one-dimensional vector through a flattening operation (Flatten). This one-dimensional feature vector undergoes the processing via a linear layer (Linear) and a LeakyReLU activation function, ultimately yielding the output $\mathbf{\tilde{S}}_r\in \mathbb{R}^{T\times d_{\text{model}}}$. $d_\text{model}$ is the feature dimension of the column vector in the matrix $\mathbf{\tilde{S}}_r$. Similarly, after the Encoding Block module, the imaginary component is transformed from $\mathbf{\bar{S}}_i\in \mathbb{R}^{T\times N\times M}$ to $\mathbf{\tilde{S}}_i\in \mathbb{R}^{T\times d_{\text{model}}}$.\par 

In the multi-head attention module, for each head $k \in \left\{ 1,\cdots ,K \right\} $ within the module, taking the real part data $\mathbf{\tilde{S}}_r$ as an example, we define the query matrix, key matrix, and value matrix as $\mathbf{Q}_{k}^{r}=\mathbf{\tilde{S}}_r\mathbf{W}_{k}^{Q}$, $\mathbf{K}_{k}^{r}=\mathbf{\tilde{S}}_r\mathbf{W}_{k}^{K}$ and $\mathbf{V}_{k}^{r}=\mathbf{\tilde{S}}_r\mathbf{W}_{k}^{V}$, respectively. The reprogramming operation in each attention head is defined as
\begin{equation}
    \label{eq:11}
    \mathbf{S}_{k}^{r}=\text{ATTENTION}\left( \mathbf{Q}_{k}^{r},\mathbf{K}_{k}^{r},\mathbf{V}_{k}^{r} \right) .
\end{equation}\par
Similarly, we also apply the $K$-head multi-head attention module to the imaginary part data $\mathbf{\bar{S}}_i$:
\begin{equation}
    \label{eq:12}
    \mathbf{S}_{k}^{i}=\text{ATTENTION}\left( \mathbf{Q}_{k}^{i},\mathbf{K}_{k}^{i},\mathbf{V}_{k}^{i} \right) ,
\end{equation}
where $\mathbf{S}_{k}^{r}$ and $\mathbf{S}_{k}^{i}\in \mathbb{R}^{T\times d}$. Subsequently, we will integrate the features derived from each head to obtain $\mathbf{S}^r$ and $\mathbf{S}^i\in \mathbb{R}^{T\times d_{\text{model}}}$. In general, we set $d=d_{\text{model}}/K$. Ultimately, we will concatenate these two data to produce the data $\mathbf{X}\in \mathbb{R}^{T\times 2\times d_{\text{model}}}$.\par
Before the data $\mathbf{X}\in \mathbb{R}^{T\times 2\times d_{\text{model}}}$ is input to the backbone network, the data needs to go through the Input Projection module for further feature extraction and dimension transformation. The processing flow of the Input Projection module is shown below:
\begin{equation}
    \label{eq:input-projection}
    \mathbf{\tilde{X}}=\text{rearrange}\left( \text{Linear}\left( \text{GELU}\left( \text{Linear}\left( \text{rearrange}\left(\mathbf{X}\right) \right) \right) \right) \right),
\end{equation}
where $\mathbf{\tilde{X}}\in \mathbb{R}^{F\times d_{\text{model}}}$. The hidden layer dimension of the two-layer Linear neural network is $d_l$. GELU (Gaussian Error Linear Unit) is the nonlinear activation function. $\text{rearrange}\left(\cdot\right)$ function implements the dimension transformation of the feature tensor.\par

\subsubsection{Backbone network}\label{backbone_network}
Recent studies have demonstrated that fine-tuned LLMs can be effectively utilized within the physical layer of wireless communication systems, yielding impressive outcomes \cite{Fan2024CsiLLMAN, LLM4CP_2024, beam_prediction_LLM_2024}. Motivated by these findings, we aim to leverage the robust modeling capabilities of the LLMs to accomplish our channel table prediction task, subsequently facilitating the port prediction for the FA.\par

The LLM selected for our study is the GPT-2 model \cite{radford2019language}. We implement LoRA fine-tuning on the pre-trained GPT-2 model. LoRA fine-tuning is based on the intrinsic low-rank characteristics of the LLMs. It simulates full parameter fine-tuning by adding bypass matrices, aiming to achieve lightweight fine-tuning \cite{Hu2021LoRALA}. In particular, we exclusively conduct LoRA fine-tuning and retraining on the query matrix $\mathbf{Q}_{k}$ and the value matrix $\mathbf{V}_{k}$ computations within the multi-head attention component of the GPT-2 model, while keeping the remaining model parameters frozen. This approach can substantially reduce the computational resources necessary for our model retraining and better utilize the knowledge acquired by GPT-2 during its pre-training phase. It is assumed that the input data for the module requiring LoRA fine-tuning is denoted by $\mathbf{Z}$. The process of LoRA fine-tuning is outlined as follows:
\begin{equation}
    \label{eq:13}
\mathbf{Q}_k=\mathbf{W}_{k}^{Q}\mathbf{Z}+\mathbf{B}_{k}^{Q}\mathbf{A}_{k}^{Q}\mathbf{Z}+\mathbf{b}_{k}^{Q},
\end{equation}
\begin{equation}
    \label{eq:14}
\mathbf{V}_k=\mathbf{W}_{k}^{V}\mathbf{Z}+\mathbf{B}_{k}^{V}\mathbf{A}_{k}^{V}\mathbf{Z}+\mathbf{b}_{k}^{V},
\end{equation}
where $\mathbf{W}_{k}^{Q}, \mathbf{W}_{k}^{V}$ are the weights of the pre-trained GPT-2 model, which remain constant and are not subject to gradient updates throughout the training process. $\mathbf{b}_k^{Q}$ and $\mathbf{b}_k^{V}$ are the biases of the loaded model, which also remain fixed. $\mathbf{B}_k^{Q}, \mathbf{B}_k^{V}\in \mathbb{R}^{d_m\times r}$ and $\mathbf{A}_k^{Q}, \mathbf{A}_k^{V}\in \mathbb{R}^{r\times d}$ are learnable parameters. Furthermore, $r\ll \min \left( d_m,d \right) $, resulting in negligible additional inference delay during model prediction when employing LoRA fine-tuning. As illustrated in Fig. \ref{fig:PORT-LLM}, we employ random Gaussian initialization for parameter $\mathbf{A}_k^{Q}$ and $\mathbf{A}_k^{V}$, and zero initialization for parameter $\mathbf{B}_k^{Q}$ and $\mathbf{B}_k^{V}$ prior to the commencement of our model training. \par
Generally, the data $\mathbf{\tilde{X}}\in \mathbb{R}^{F\times d_{\text{model}}}$ is integrated into the backbone network, where the following procedure occurs:
\begin{equation}
    \label{eq:16}
    \mathbf{X}_{\text{LLM}}=\text{LLM}_{\text{LoRA}}\left( \mathbf{\tilde{X}} \right) \in \mathbb{R}^{F\times d_{\text{model}}},
\end{equation}
where $\text{LLM}_{\text{LoRA}}\left( \cdot \right) $ represents the LLM-based backbone network that has been fine-tuned by LoRA.\par

\subsubsection{Output Projection}\label{output_projection}
In the Output Projection module, a two-layer linear neural network is employed in conjunction with the rearrange operation to derive the final output of the model in the following manner:
\begin{equation}
    \label{eq:output-projection}
    \mathbf{Y}=\text{rearrange}\left( \text{Linear}\left( \text{GELU}\left( \text{Linear}\left( \text{rearrange}\left( \mathbf{X}_\text{LLM} \right) \right) \right) \right) \right),
\end{equation}
where $\mathbf{Y} \in \mathbb{R}^{F\times 2\times N\times M}$.\par
% \begin{equation}
%     \label{eq:17}
%     \mathbf{Y}=\text{rearrange}\left( \text{FC}\left( \mathbf{X}_{\text{LLM}} \right) \right) \in \mathbb{R}^{F\times 2\times N\times M}.
% \end{equation}\par
Subsequently, execute the denormalization process
\begin{equation}
    \label{eq:18}
    \mathbf{\hat{Y}}=\mathbf{\sigma Y}+\mathbf{\mu },
\end{equation}
where $\mathbf{\hat{Y}}\in \mathbb{R}^{F\times 2\times N\times M}$. The second dimension corresponds to the real and imaginary components of the prediction channel tables, respectively. Additionally, the final output data 
$\mathbf{\hat{S}}\in \mathbb{C}^{F\times N\times M}$ is obtained as follows:
\begin{equation}
    \label{eq:output-data}
    \mathbf{\hat{S}}=\mathbf{\hat{Y}}\left[ :,0,:,: \right] +j\mathbf{\hat{Y}}\left[ :,1,:,: \right]. 
\end{equation}\par

\begin{figure*}[t]
    \centering
    \includegraphics[scale=0.95]{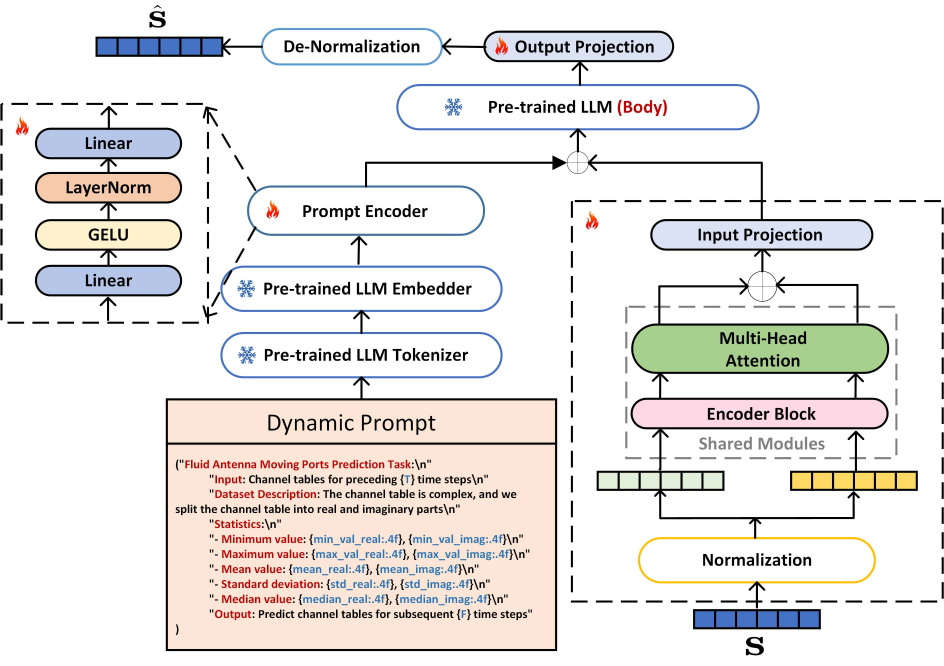}
    \caption{The architecture of our proposed Prompt-Port-LLM model.}
    \label{fig:PORT-LLM-prompt}
\end{figure*}

\subsubsection{Moving Port prediction}\label{predicted-ports}
Upon acquiring the channel tables $\mathbf{\hat{S}}=\left\{ \mathbf{\hat{S}}_1,\mathbf{\hat{S}}_2,\cdots ,\mathbf{\hat{S}}_F \right\} \in \mathbb{C}^{F\times N\times M}$ for the subsequent $F$ moments, we employ the predicted channel tables in conjunction with the associated known reference channel $\mathbf{H}_{\text{ref}}=\left\{ \mathbf{H}_{\text{ref}_1},\mathbf{H}_{\text{ref}_2},\cdots ,\mathbf{H}_{\text{ref}_F} \right\} \in \mathbb{C}^{F\times N\times M} $ to derive the final predicted moving ports of the FA for the forthcoming $F$ moments, utilizing the following formula:
\begin{equation}
    \label{eq:optimal-port}
    \mathbf{p}_i=\text{unravel\_index}\left( \text{argmin}\left( \left| \mathbf{\hat{S}}_i-\mathbf{H}_{\text{ref}_i} \right| \right) ,\left( N,M \right) \right)  ,
\end{equation}
where $\mathbf{P}=\left[ \mathbf{p}_1,\cdots ,\mathbf{p}_F \right] \in \mathbb{R}^{F\times 2\times 1}, \mathbf{p}_i=\left[ n_i,m_i \right] ^T\in \mathbb{R}^{2\times 1} , 1\le i\le F, 1\le n_i\le N,1\le m_i\le M$ denotes the predicted moving port of FA at the subsequent $i$-th moment. Here, $n_i$ and $m_i$ correspond to the port indices associated with the predicted moving port of FA along the $z$-axis and $y$-axis, respectively, at the subsequent $i$-th time instance.\par

\subsection{The Port-LLM model with prompt fine-tuning}

The Port-LLM model that utilizes prompt fine-tuning, referred to as Prompt-Port-LLM, shares a similar network architecture with the original Port-LLM model. The primary distinction between the two models lies in their fine-tuning strategies. Specifically, the original Port-LLM employs the LoRA fine-tuning approach, while the Prompt-Port-LLM incorporates the prompt fine-tuning mechanism. Furthermore, we design a specialized Prompt Encoder module to make prompts trainable.\par

As illustrated in Fig. 4, the dynamic prompts information associated with the Prompt-Port-LLM model encompasses the description of the model's task, the characteristics of the dataset, and the dynamic statistical properties of the input data throughout the training process. These dynamic statistical properties are computed in real time and include the maximum and minimum values, mean, standard deviation, and median of the input data.\par

Dynamic prompts need to undergo processing by the tokenizer and embedder components of the loaded pre-trained LLM. The pre-trained LLM tokenizer transforms the prompt text into a discrete sequence of symbolic tokens, while the pre-trained LLM embedder subsequently converts this sequence into a dense vector representation suitable for neural network processing. During the prompt encoding phase, we design a specialized Prompt Encoder module to deeply encode the prompt vectors. This module employs a two-layer fully connected (linear) network architecture, with the hidden layer dimension set to $d_\text{model}$, which facilitates dynamic optimization of the prompt through the utilization of trainable parameters.\par

\subsection{Optimization objectives}\label{subsec_optimation_objection}
\begin{figure}[h]
    \centering
    \includegraphics[scale=0.30]{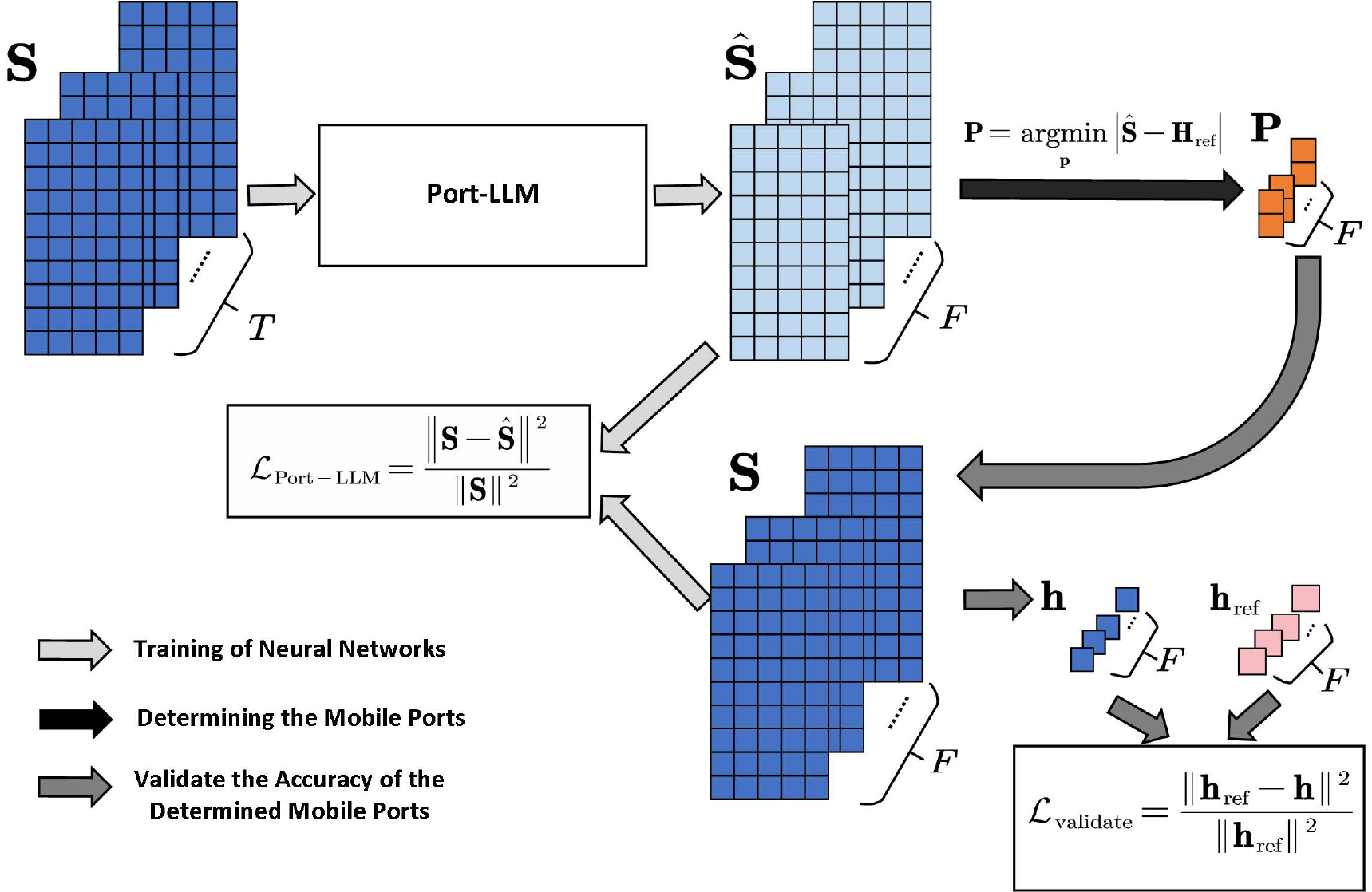}
    \caption{The flowchart for predicting the moving ports of the FA based on our proposed model.}
    \label{fig:流程图}
\end{figure}
Our proposed model is initially trained on channel table datasets and then applied for testing. During the model training process, the objective function is the normalized mean square error (NMSE) between the channel tables $\mathbf{\hat{S}}$ predicted by our model for the future $F$ moments and the actual channel tables $\mathbf{S}$ for these $F$ moments.\par
\begin{equation}
    \label{eq:18}
    \mathcal{L}_{\text{Port}-\text{LLM}}=\frac{\lVert \mathbf{S}-\mathbf{\hat{S}} \rVert ^2}{\lVert \mathbf{S} \rVert ^2}.
\end{equation}\par

\begin{table}[t!]\normalsize
    \centering
    \begin{tabular}{p{8cm}} %设置表格两列内容为左对齐
    \toprule %顶线
    \textbf{Algorithm 1:} {A Port Prediction Method for Fluid Antenna based on the Proposed Port-LLM} \\ 
    \midrule %中线
    \textbf{Input:} {The channel tables at the last $T$ moments $\mathbf{S}$, the corresponding reference channel tables $\mathbf{H}_{\text{ref}}$}\\
    \textbf{Output:} {Predicted moving ports of FA at subsequent $F$ moments $\mathbf{P}$}\\
    \textbf{Initialization:} {Learning rate of our model $\alpha_{\text{Port-LLM}}$, exponential decay rate of moment estimates $\beta _{\text{Port-LLM}}$, batch size of the train sets $m_1$, batch size of the test sets $m_2$, the number of epochs $\mathcal{K}$}\\
    \textbf{Process:}\\
    \textbf{For} $\text{epoch}=1,2,\cdots ,\mathcal{K}$ \textbf{do}\\
    \begin{itemize}
    \item 
    %Utilize the channel tables data at the last $T$ moments $\mathbf{S}$ as the input of our proposed model to 
    Forecast the channel tables for the future $F$ time intervals $\mathbf{\hat{S}}$:
        \begin{equation}
            \mathbf{\hat{S}}=\text{Net}_{\text{Port}-\text{LLM}}\left( \mathbf{S} \right). \nonumber
        \end{equation}
    \item Update our proposed model by using adaptive moment estimation:
          \begin{equation}
              \mathcal{L}_{\text{Port}-\text{LLM}}=\frac{\lVert \mathbf{S}-\mathbf{\hat{S}} \rVert ^2}{\lVert \mathbf{S} \rVert ^2}. \nonumber
          \end{equation}
    \item Obtain the predicted moving ports of FA $\mathbf{P}=\left[ \mathbf{p}_1,\cdots ,\mathbf{p}_F \right]$ for the subsequent $F$ moments:
            \begin{equation}
             \begin{aligned}
                \mathbf{p}_i = \text{unravel\_index}\bigg( 
                &\text{argmin} \left( \left| \mathbf{\hat{S}}_i - \mathbf{H}_{\text{ref}_i} \right| \right), \\
                &\!\left( N, M \right) \bigg)
                \end{aligned}.  \nonumber
            \end{equation}
    \item Obtain the corresponding port channel $\mathbf{h}=\left[ h_1,h_2,\cdots ,h_F \right] ^T$ in the actual channel tables based on the predicted ports.
    \item Validate the accuracy of the predicted ports by our model, and calculate the NMSE between the channel of the port predicted by our model and the reference channel:
    %Validate the accuracy of the predicted ports by our model:
    %, and compute the error between the channel corresponding to the predicted ports and the reference channel:
            \begin{equation}
                \mathcal{L}_{\text{validate}}=\frac{\lVert \mathbf{h}_{\text{ref}}-\mathbf{h} \rVert ^2}{\lVert \mathbf{h} \rVert ^2}.\nonumber
            \end{equation}
    \end{itemize}
    \textbf{End for}\\
    \bottomrule %底线
    \end{tabular}
\end{table}

The primary objective of this study is to forecast the moving ports of the FA for future time intervals. Attaining this objective necessitates the completion of two distinct phases. Firstly, we will employ our proposed neural network model to predict the channel tables for the forthcoming $F$ time intervals, utilizing the channel tables from the preceding $T$ time intervals as input. Subsequently, we will employ Eq. (\ref{eq:optimal-port}) to calculate the moving port of the FA corresponding to the specified future $F$ time intervals. The comprehensive implementation procedure is illustrated in Fig. \ref{fig:流程图}. The pseudocode for the algorithm predicting moving ports of the FA for future moments based on Port-LLM is outlined in \textbf{Algorithm 1}. Note that the algorithmic flow of our proposed Prompt-Port-LLM model for predicting the moving port of FA is consistent with that of the Port-LLM model.\par

\section{Numerical Results}\label{sec:numericalResult}
This section will outline the simulation settings utilized for our model, assess its performance across various evaluation metrics, and conduct the comparative analysis with established methodologies for addressing FA moving ports.\par

Note that during our proposed model training and performance testing, to reduce the number of model parameters, we employ single-antenna data in the single-input single-output (SISO) setting for model training. Nevertheless, given that practical BSs are typically equipped with multiple antennas, our trained model is directly utilized for testing in the multiple-input single-output (MISO) scenario without undergoing any retraining.\par

\subsection{Simulation Settings}\label{subsec_experimental_settings}

\begin{table}[t!]
    \centering
    \caption{The Main Simulation Parameters}
    \label{tab:simulation-parameters}
    \begin{tabularx}{\linewidth}{|>{\centering\arraybackslash}c|>{\centering\arraybackslash}X|}
        \hline
        Channel Model & CDL-D \\ \hline
        Carrier Frequency (GHz)   & 39   \\ \hline
        CSI Delay (ms) & 4 
        \\ \hline
        Delay Spread (ns) & 616 
        \\ \hline
        Sampling Time  & $T_0=5, 6, 10$
        \\ \hline
        UE FA Configuration & 
        \begin{minipage}{\linewidth}
        \centering
        \footnotesize
        $\left(W_y,W_z\right)=\left(10,20\right)$,\\
        $\left(M,N\right)=\left(100,50\right)$,\\
        $\left(\rho_y,\rho_z\right)=\left(5,5\right)$
        \end{minipage}
        \\ \hline
        RMS Angular Spreads       & 
        \begin{minipage}{\linewidth}
        \centering
        \footnotesize
        $\left[31^\degree, 149^\degree, 150^\degree, 30^\degree\right]$, $\left[-38^\degree, 218^\degree, 227^\degree, -47^\degree\right]$, \\
        $\left[1^\degree, 179^\degree, 99^\degree, 81^\degree\right]$, $\left[10^\degree, 170^\degree, 36^\degree, 144^\degree\right]$, \\
        $\left[149^\degree, 31^\degree, 53^\degree, 127^\degree\right]$, $\left[129^\degree, 51\degree, 71^\degree, 109^\degree\right]$, \\
        $\left[-15^\degree, 195^\degree, 210^\degree, -30^\degree\right]$, $\left[199^\degree, -19^\degree, 212^\degree, -32^\degree\right]$, \\
        $\left[-43^\degree, 223^\degree, 76^\degree, 104^\degree\right]$, $\left[7^\degree, 173^\degree, 23^\degree, 157^\degree\right]$
        \end{minipage}
        \\ \hline
    \end{tabularx}
\end{table}

\subsubsection{Dataset}
To mitigate the computational demands associated with the training of our model, we adopt a SISO system for the acquisition of training datasets. In this configuration, the antenna on the BS side remains stationary, while the UE side is equipped with the FA. This FA can move in a two-dimensional plane of dimensions $10\lambda \times 20\lambda$, situated within the $y$-$z$ plane. The quantities of movable antenna ports along the $y$-axis and $z$-axis are $M=50$ and $N=100$, respectively. The densities of ports along the $y$-axis and $z$-axis are $\rho_y=\rho_z=5$. The carrier frequency $f$ utilized in this study is 39 GHz, and we employ the clustered delay line (CDL) channel model as defined by the 3rd generation partnership project (3GPP) \cite{3GPP2019Study}. The channel model includes 37 paths, which comprise a LoS path and 36 NLoS paths. The velocity of UEs ranges from 90 km/h to 150 km/h. Each slot contains 14 OFDM symbols, and the duration of a slot is 1 ms. Each group of 50 time slots has a sampling time. The channel corresponding to this sampling moment $T_0$ serves as the reference channel for that group time. Furthermore, the reference channel is accessible to the UE. To enhance the quantity and diversity of the training dataset, we conduct simulations of communication channels for 10 UEs positioned in various orientations. During the simulation of each UE, we randomly select a distinct sampling time $T_0$ within the designated time interval. The specific values of the Root Mean Square (RMS) angular spreads of AOD, EOD, AOA, and EOA for these 10 UEs are shown in Table \ref{tab:simulation-parameters}. A total of 54,300 samples are collected, with 75\% of the dataset allocated for the training set and the remaining 25\% designated for the test set.\par

\subsubsection{Network and Training Parameters}\label{traing_setup}
In the simulation of our model, the forecasting period for the FA moving ports is established at $F=8$. Concurrently, the duration for the employed channel tables is also designated as $T=8$. As previously indicated, the dimension of the FA moving port table is set as $N\times M=100\times 50$. We employ the smallest version of the GPT-2 model with 768 feature dimensions and utilize only the initial $N_L=6$ layers of the pre-trained GPT-2 architecture. Furthermore, the number of heads of the multi-head attention module employed in our model is $K=8$. And the dimension is set as $d_{\text{model}}=768$ in the multi-head attention. In the input projection module, the hidden layer dimension $d_l$ of the two-layer Linear neural network is set as 2048. Furthermore, in the proposed Prompt-Port-LLM model, the hidden layer dimension of the prompt encoder module is specified as 768. The Adam algorithm is employed to update the parameters. The specific simulation parameters of our model can be found in Table \ref{tab:net-parameters}. In the training process of our proposed model with the warm-up aided cosine learning rate (LR) schedule, the learning rate of our model $\alpha_{\text{Port-LLM}}$ undergoes a linear increase from $\alpha_{\min}=4\times10^{-6}$ to $\alpha_{\max}=1\times 10^{-3}$ in the initial 100 epochs, known as ``warm-up", as described in Eq. (\ref{eq:warm-up}).
\begin{equation}
    \label{eq:warm-up}
    \alpha_{\text{Port-LLM}} =\alpha _{\min}+\left( \alpha _{\max}-\alpha _{\min} \right) \cdot \frac{t}{T_{\max}},
\end{equation}
where $T_{\max} = 100$ and $t$ are the number of total warm-up epochs and current number of training epoch, respectively.\par
In the ``cosine decay" phase, the LR of the model is cosine decaying. In general, the LR of our model varies as follows during our model training:\par
\begin{equation}
    \label{eq:lr}
    \alpha_{\text{Port-LLM}} =\alpha _{\min}+\frac{1}{2}\left( \alpha _{\max}-\alpha _{\min} \right) \left( 1+\cos \left( \frac{t-T_{\max}}{T-T_{\max}}\pi \right) \right),
\end{equation}
This warm-up-aided cosine annealing algorithm facilitates rapid convergence of our model in the early stages and prevents it from being stuck in local optima due to high learning rates in later stages.\par

\begin{table}[t!]\normalsize
    \centering
    \caption{Hyper-parameters for network training}
    \begin{tabular}{C{5cm}|C{1.2cm}|C{1.2cm}}
    \hline
         \multicolumn{2}{c|}{\textbf{Port-LLM Parameters}} & \textbf{Value}  \\ 
         \hline
          \multirow{2}{*}{Learning rate}   & $\alpha_{\text{max}}$ & $1\times 10^{-3}$  \\
         \cline{2-3}
         & $\alpha_{\text{min}}$ & $4\times 10^{-6}$ \\
         \hline
          Exponential decay rate & $\beta_{\text{Port-LLM}}$ & $\left(0.9, 0.99\right)$ \\
         \hline
          \multirow{2}{*}{Batch size} & $m_1$ & 200 \\
         \cline{2-3}
         & $m_2$ & 200 \\
         \hline
         Number of epochs & $\mathcal{K}$& 600\\
         \hline
    \end{tabular}
    \label{tab:net-parameters}
\end{table}

\subsubsection{Baselines}\label{baselines}
To assess the efficacy of our proposed model, we conduct a comparative analysis of various model-based and deep learning-based methods for FA moving port calculations, which served as benchmarks.\par
\begin{itemize}
    \item \textbf{MPMP} \cite{Li2024TransformingTT}: MPMP is a model-based methodology that employs the FA to tackle challenges associated with mobility, utilizing a matrix pencil approach for predicting mobility ports. In the comparative experiment, the mobility port prediction technique based on MPMP has a one-dimensional mobility area for the FA. This mobility region is oriented along the $z$-axis, measuring $20\lambda$ in size and encompassing a total of 100 ports.\par

    \item \textbf{Vec Prony} \cite{yin2020addressing}: The Vector Prony-based channel prediction algorithm is also a model-based method. In this approach, a second-order Vec Prony algorithm is utilized.\par

    \item \textbf{RNN} \cite{Lipton2015ACR}: RNN is a conventional neural network architecture designed for the analysis of sequential data. In our study, we substitute the pre-trained GPT-2 model integrated into our proposed model with an RNN model. The experimental setup involve the utilization of a two-layer RNN network.\par

    \item \textbf{LSTM} \cite{graves2012long}: LSTM is a specialized form of RNNs that effectively mitigate the issues of gradient vanishing and explosion that are commonly faced by conventional RNNs when handling extended sequences. 
    %This is achieved through the implementation of meticulously designed gating mechanisms and cell states, which enable LSTM to better capture long-term dependencies that traditional RNNs struggle to manage. 
    In our experiment, we employ a two-layer LSTM model as a substitute for the loaded GPT-2 model integrated within our framework.\par

    \item \textbf{GRU} \cite{Chung2014EmpiricalEO}: GRU is a streamlined adaptation of the LSTM architecture. It is engineered to maintain the capacity of LSTM for managing long-term dependencies while simultaneously decreasing computational complexity and the total number of parameters within the model. Similarly, we implement a two-layer GRU model as an alternative to the loaded GPT-2 model within our framework.\par
    
    \item \textbf{Transformer} \cite{transformer}: The Transformer is a deep learning model architecture that departs from conventional recurrent neural network frameworks, including RNN, LSTM, and GRU, by utilizing the attention mechanism exclusively for the processing of sequential data. In experiment, we employ a Transformer model as a substitute for the loaded GPT-2 model within our framework. Specifically, this Transformer model consists of an 8-head multi-head attention module, with an input dimension of 768 and an embedding dimension of 512. In addition, the hidden layer dimension of the multilayer perceptron (MLP) inside the Transformer model is set to 3072.\par
    
\end{itemize}

\subsubsection{Performance Metrics}\label{per_indices}
Since our proposed Port-LLM-based model predicts the moving ports of the FA in two steps, the first step is to predict the channel tables, and the second step is to obtain the future moving ports of the FA based on the predicted channel tables. Therefore, during the model training process, we set several metrics to evaluate the performance of our proposed models. Specifically, $\text{Accuracy}$ is employed to measure the precision of the model in predicting the channel table, as described in Eq. (34).\par

\begin{equation}
    \label{eq:acc}
    \text{Accuracy}=\left( 1-\frac{\left| \mathbf{\hat{S}}-\mathbf{S} \right|}{\left| \mathbf{S} \right|} \right) \times 100\%.
\end{equation}\par
$\text{NMSE}_{\text{v}}$ is used to evaluate the NMSE between the channel corresponding to the predicted moving port based on the predicted channel table of the proposed model and the reference channel. This metric is defined in Eq. (35).\par
\begin{equation}
    \label{eq:validate_nmse}
    \text{NMSE}_{\text{v}}=10\log _{10}\left\{ \mathbb{E}\left[ \frac{\lVert \mathbf{h}-\mathbf{h}_{\text{ref}} \rVert ^2}{\lVert \mathbf{h}_{\text{ref}} \rVert ^2} \right] \right\} \left( \text{dB} \right).
\end{equation}\par
Additionally, we conduct a comparative analysis of the spectral efficiency (SE) derived from our model against the SE achieved through the Vec Prony algorithm and the MPMP algorithm, respectively. It is computed as
\begin{equation}
    \label{eq:se}
    \text{SE}=\sum_{u=1}^{N_{\text{UE}}}{\mathbb{E}}\left\{ \log _2\left( 1+\text{SINR}_u \right) \right\} \ \left( \text{bps/Hz} \right) ,
\end{equation}
where $N_{\text{UE}}$ is the number of UEs, $\text{SINR}_u$ denotes signal-to-interference-and-noise radio of the $u$-th UE. In the process of simulating and evaluating the SE of the system, we construct a DL multiuser MISO communication scenario containing 10 UEs. The simulation parameters are set as follows: signal-to-noise ratio (SNR) is tested from 0 dB to 30 dB, and the DL precoder is eigen zero-forcing (EZF) \cite{EZF_2010}.\par

\begin{figure}[t!]
    \centering
    \includegraphics[scale=0.41]{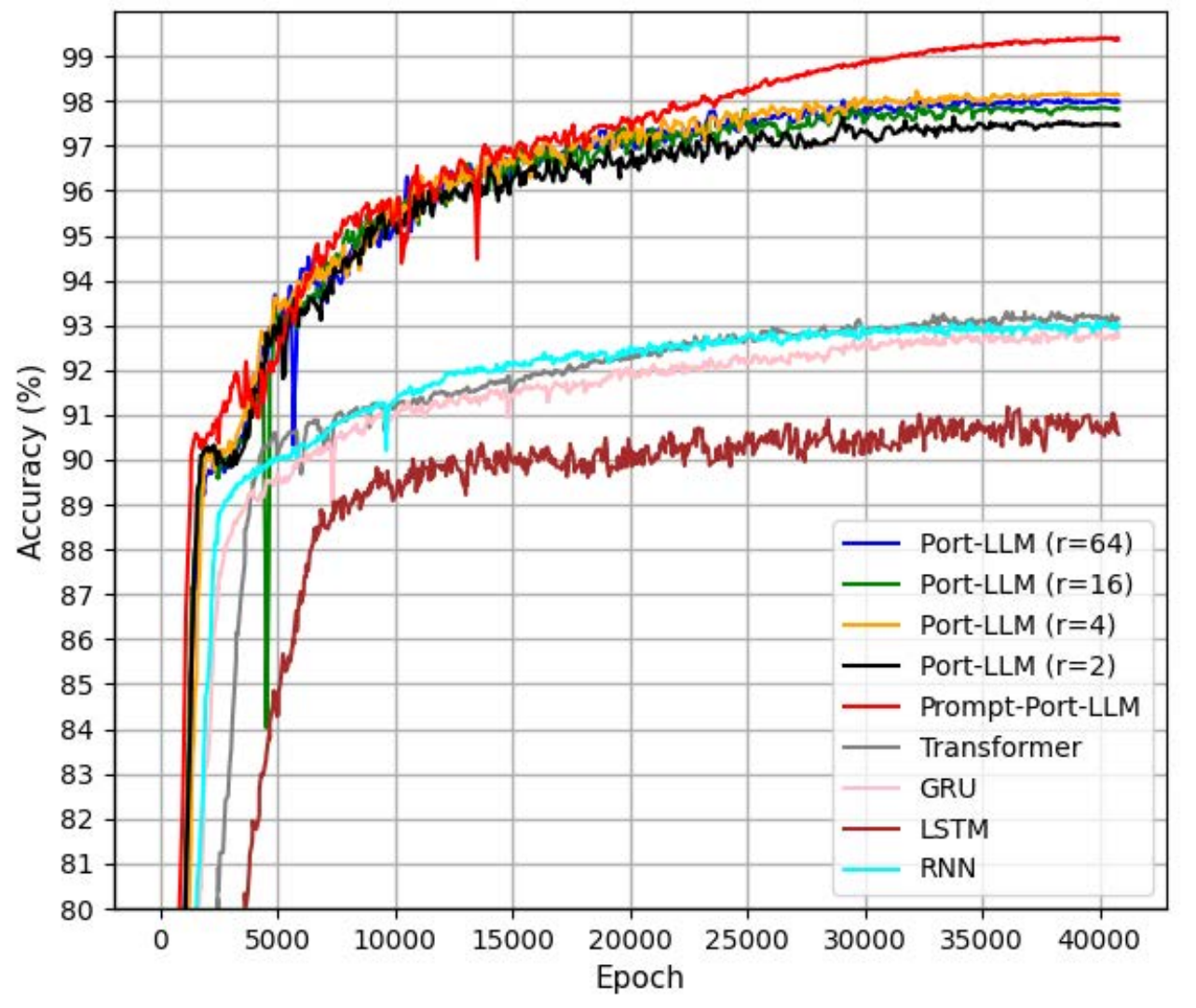}
    \caption{The prediction accuracy of test datasets during model training.}
    \label{fig:acc_test}
\end{figure}

\begin{figure}[t!]
    \centering
    \includegraphics[scale=0.37]{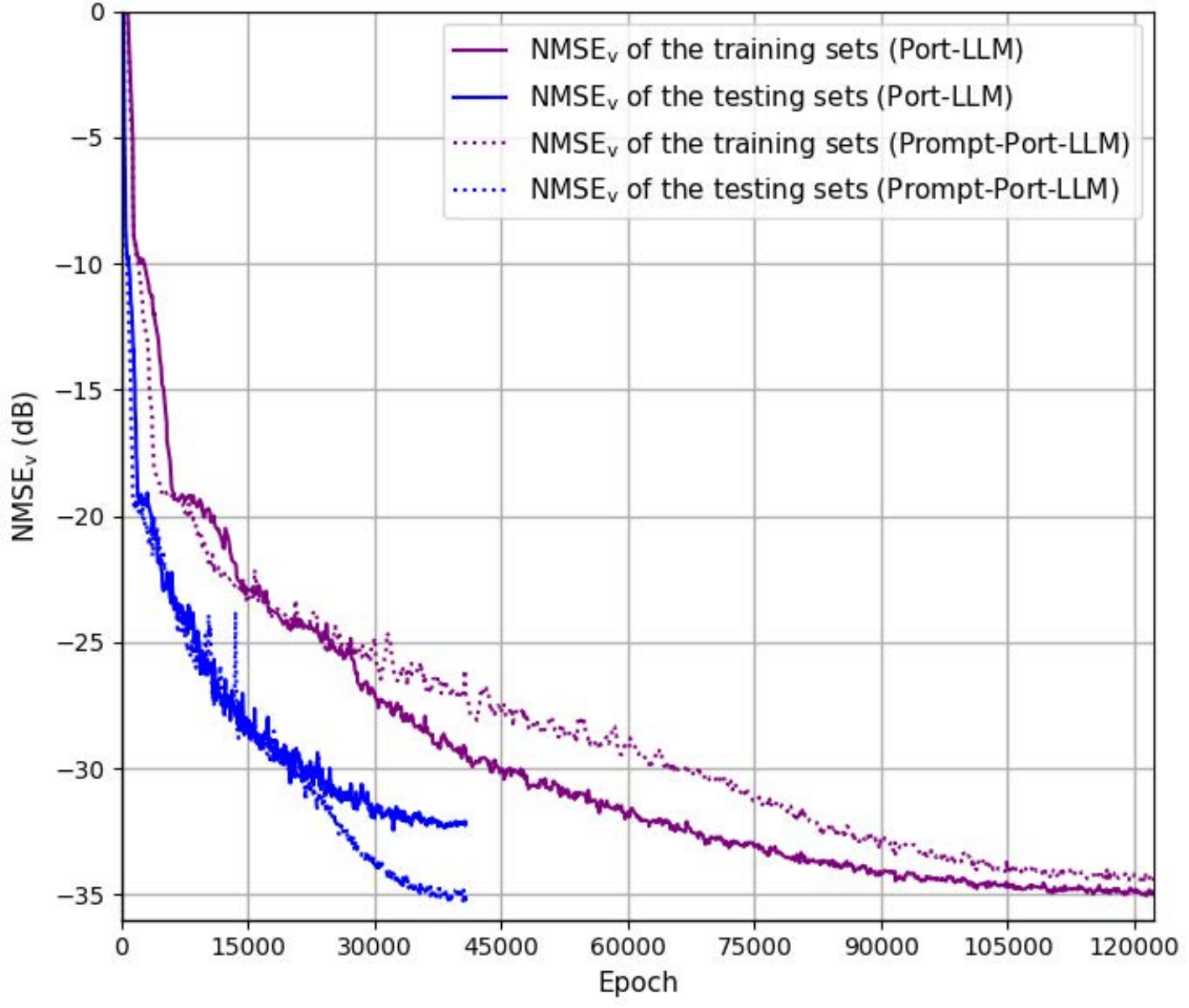}
    \caption{The NMSE of the proposed Port-LLM model vs. the number of epochs.}
    \label{fig:nmse}
\end{figure}

\begin{figure*}[t!]
\centering
\subfigure[]
{
    \begin{minipage}[b]{.31\linewidth}
        \centering
        \includegraphics[scale=0.40]{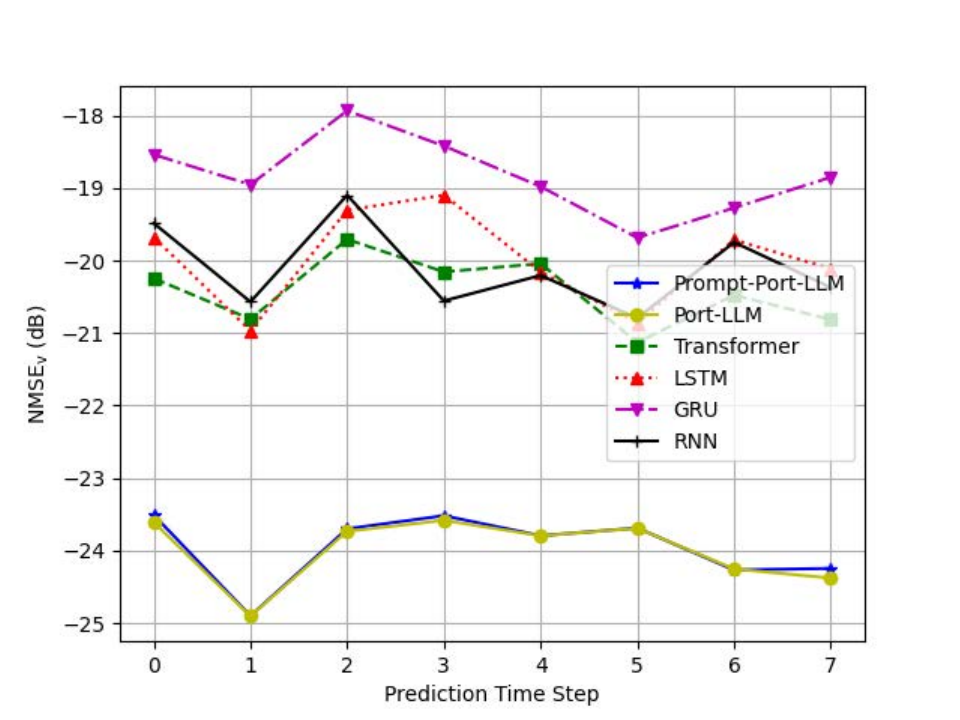}
    \end{minipage}
}
\subfigure[]
{
 	\begin{minipage}[b]{.31\linewidth}
        \centering
        \includegraphics[scale=0.40]{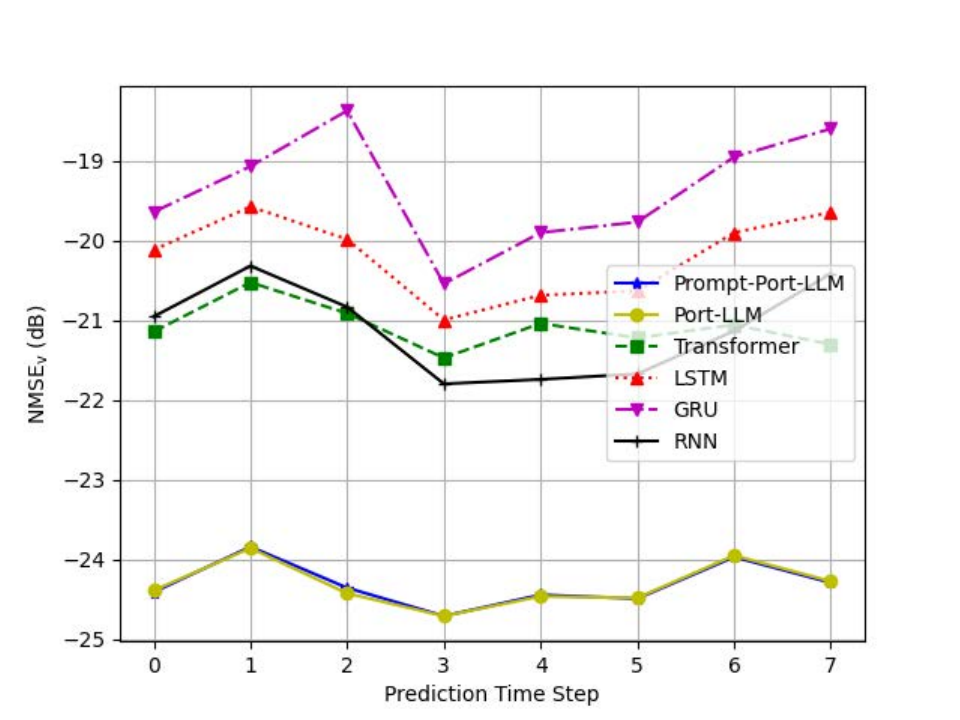}
    \end{minipage}
}
\subfigure[]
{
 	\begin{minipage}[b]{.31\linewidth}
        \centering
        \includegraphics[scale=0.40]{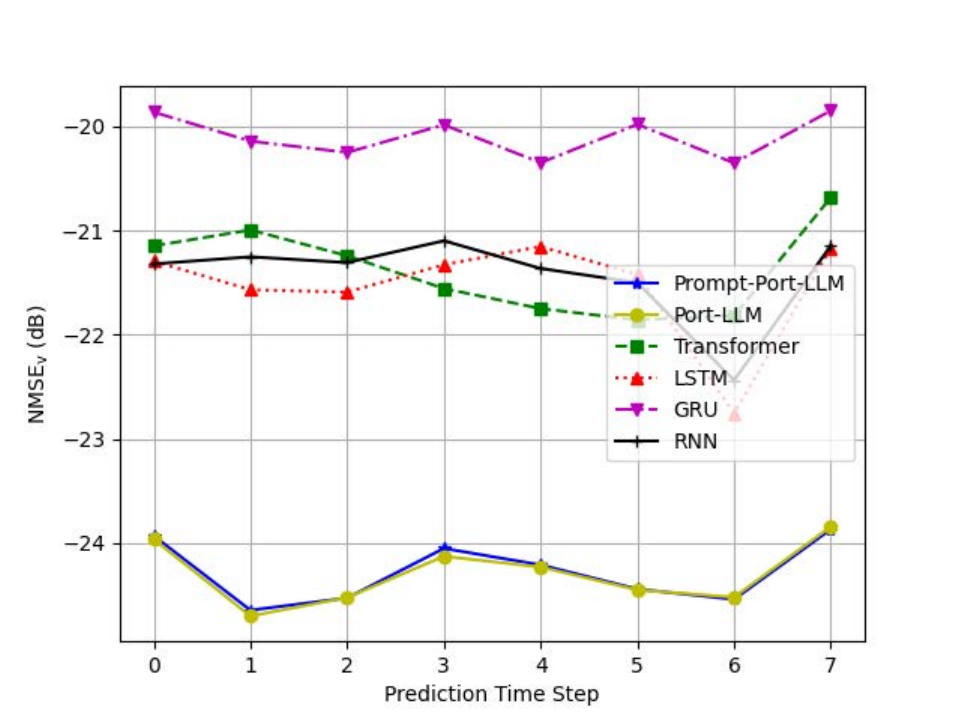}
    \end{minipage}
}
\caption{When the number of antennas on the BS side is $2\times 8$, the performance of different models under various velocities. (a) The test velocity is 90 km/h; (b) The test velocity is 120 km/h; (c) The test velocity is 150 km/h.}
\label{fig:different_models_2_8}
\end{figure*}
\begin{figure*}[t!]
\centering
\subfigure[]
{
    \begin{minipage}[b]{.31\linewidth}
        \centering
        \includegraphics[scale=0.40]{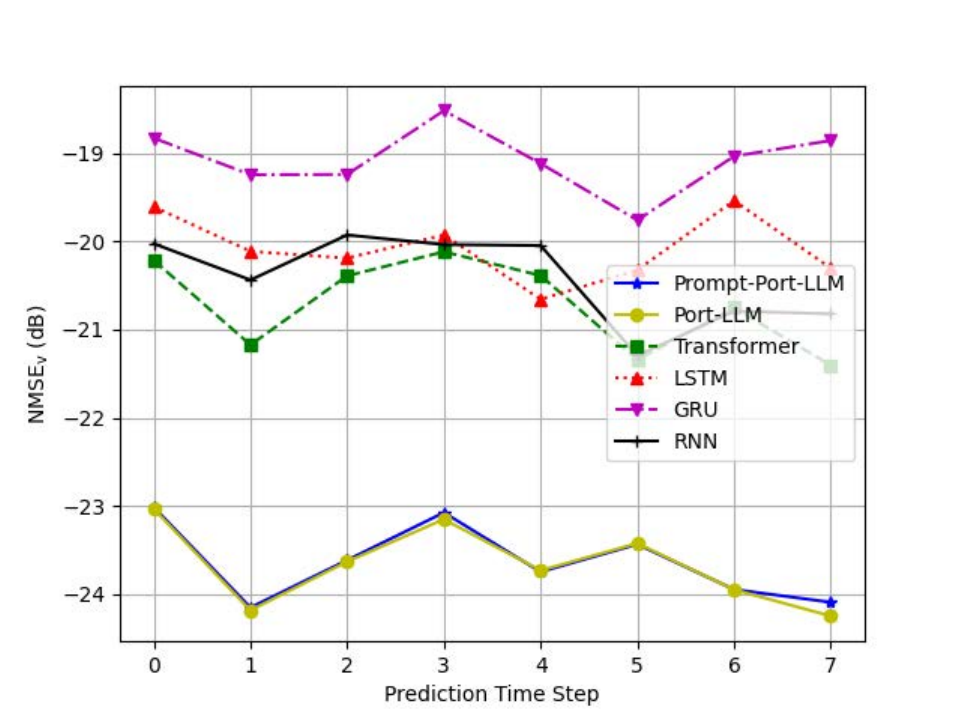}
    \end{minipage}
}
\subfigure[]
{
 	\begin{minipage}[b]{.31\linewidth}
        \centering
        \includegraphics[scale=0.40]{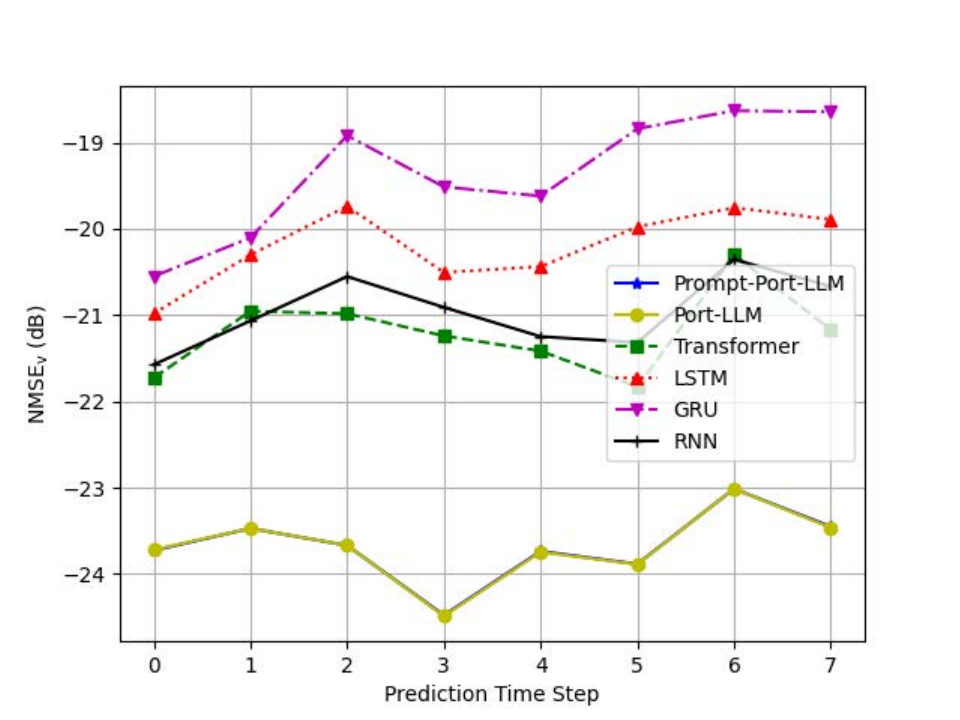}
    \end{minipage}
}
\subfigure[]
{
 	\begin{minipage}[b]{.31\linewidth}
        \centering
        \includegraphics[scale=0.40]{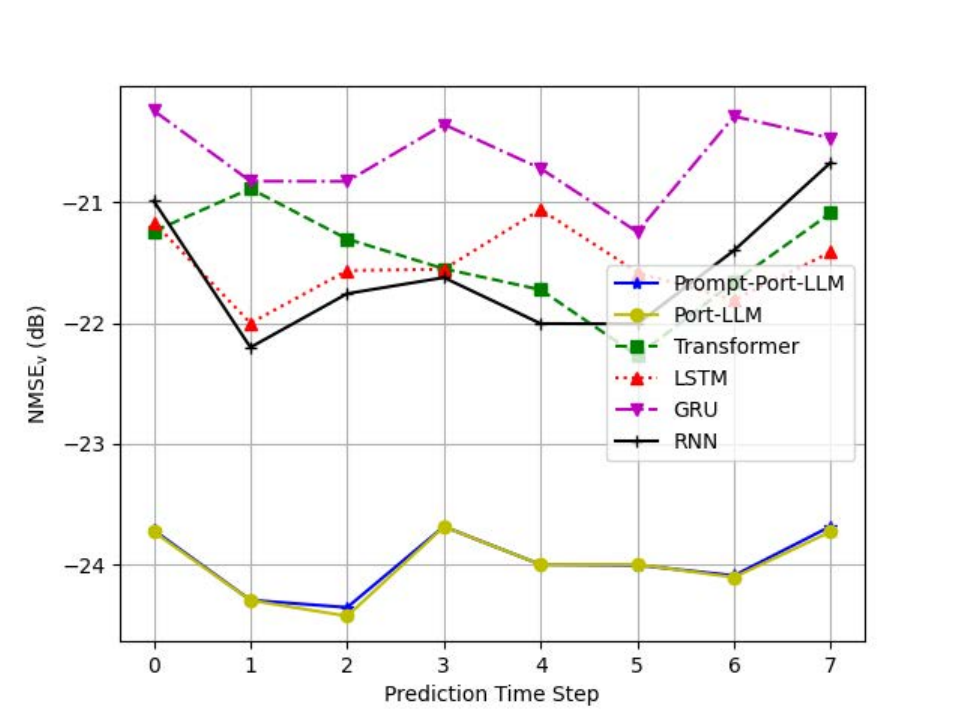}
    \end{minipage}
}
\caption{When the number of antennas on the BS side is $8\times 8$, the performance of different models under various velocities. (a) The test velocity is 90 km/h; (b) The test velocity is 120 km/h; (c) The test velocity is 150 km/h.}
\label{fig:different_models_8_8}
\end{figure*}

\subsection{Performance Evaluation}\label{subsec_performance_evaluation}
Fig. 6 shows the curve of the accuracy of the channel table predicted by various models in relation to the number of training epochs throughout the training process. For our proposed Port-LLM model, we compare the prediction performance of the model at different LoRA fine-tuning ranks $r$. The results indicate that the Port-LLM model achieves the highest accuracy when $r$ is set to 4. This phenomenon may be attributed to the fact that, within the context of our model task, selecting an excessively low LoRA fine-tuning rank $r$ yields an inadequate quantity of trainable parameters, thereby increasing the risk of underfitting. In contrast, choosing an excessively high rank $r$ results in a surplus of trainable parameters, which may result in overfitting and increase the likelihood that the model becomes trapped in local optima. Therefore, all subsequent references to the Port-LLM model refer to the model under the condition of LoRA fine-tuning with rank $r=4$. From the figure, it can be seen that the predictive accuracies of both the Port-LLM model and the Prompt-Port-LLM model surpass those of other models based on RNN, LSTM, GRU, and Transformer architectures. Furthermore, the predictive accuracy of the proposed Prompt-Port-LLM model is slightly higher than that of the proposed Port-LLM model. Specifically, the Port-LLM model attains a prediction accuracy of approximately 98.18$\%$, while the Prompt-Port-LLM model achieves an accuracy of around 99.45$\%$.\par

It should be further noted that the RNN-based, LSTM-based, GRU-based, and Transformer-based architectures have experienced gradient explosion during the training process. To mitigate this problem, we incorporate a LayerNorm layer and a dropout operation within the output projection module of these models. In contrast, our proposed Port-LLM and Prompt-Port-LLM models, which are based on LLMs, do not suffer from such problems. This observation further highlights the superiority of our proposed architecture based on the pre-trained GPT-2 model.\par

Fig. 7 illustrates the curves of the $\text{NMSE}_{\text{v}}$ between the reference channel and the channel corresponding to the moving port predicted by our proposed Port-LLM model and Prompt-Port-LLM model with the number of training epochs on both the training set and the test set during the training process. As can be seen in Fig. 7, the NMSE between the reference channel and the channel corresponding to the moving port predicted by both models ultimately converges to slightly below -30 dB. This result further shows that our proposed Port-LLM model and Prompt-Port-LLM model exhibit high accuracy in predicting the moving ports of FA at future moments.\par

In practical applications addressing mobility challenges, it is common for BS antennas to be configured as multi-antenna systems. In order to verify the effectiveness of our models in the MISO system, we investigate the performance of our models at $2\times 8$, $8\times 8$, and $32\times 8$ antenna configurations at the BS-side. Additionally, we compare the prediction performance and robustness of our proposed models against other neural network-based models, taking into account different configurations of BS antennas and varying UE mobility speeds. It is noteworthy that when evaluating the performance of our proposed models in MISO scenarios, we directly employ the previously trained model under SISO conditions without retraining for different MISO configurations. \par
\begin{figure*}[t!]
\centering
\subfigure[]
{
    \begin{minipage}[b]{.31\linewidth}
        \centering
        \includegraphics[scale=0.40]{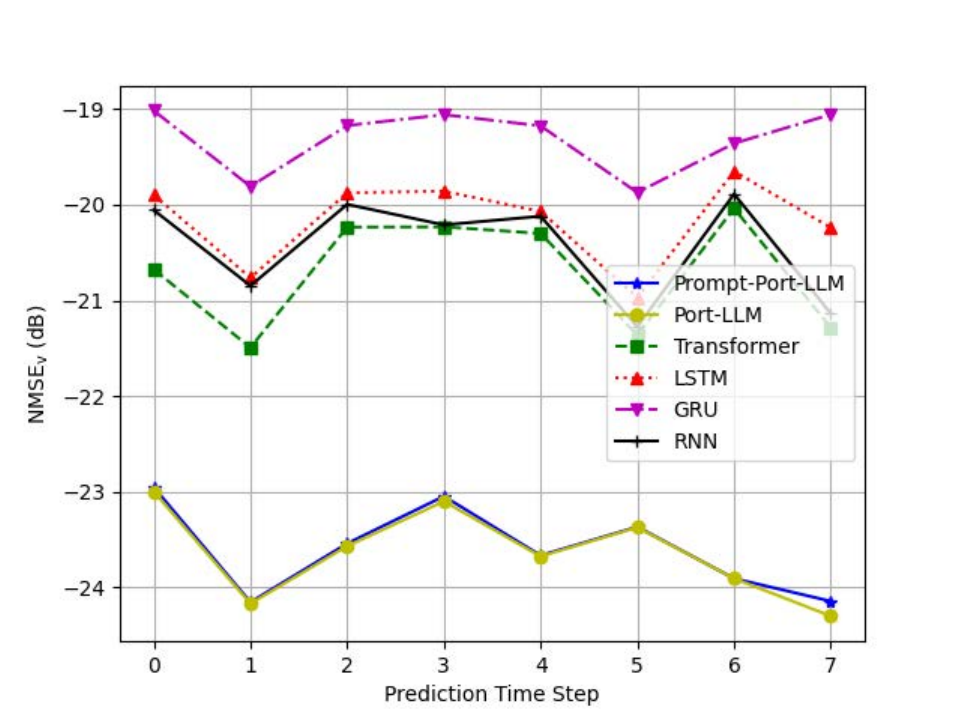}
    \end{minipage}
}
\subfigure[]
{
 	\begin{minipage}[b]{.31\linewidth}
        \centering
        \includegraphics[scale=0.40]{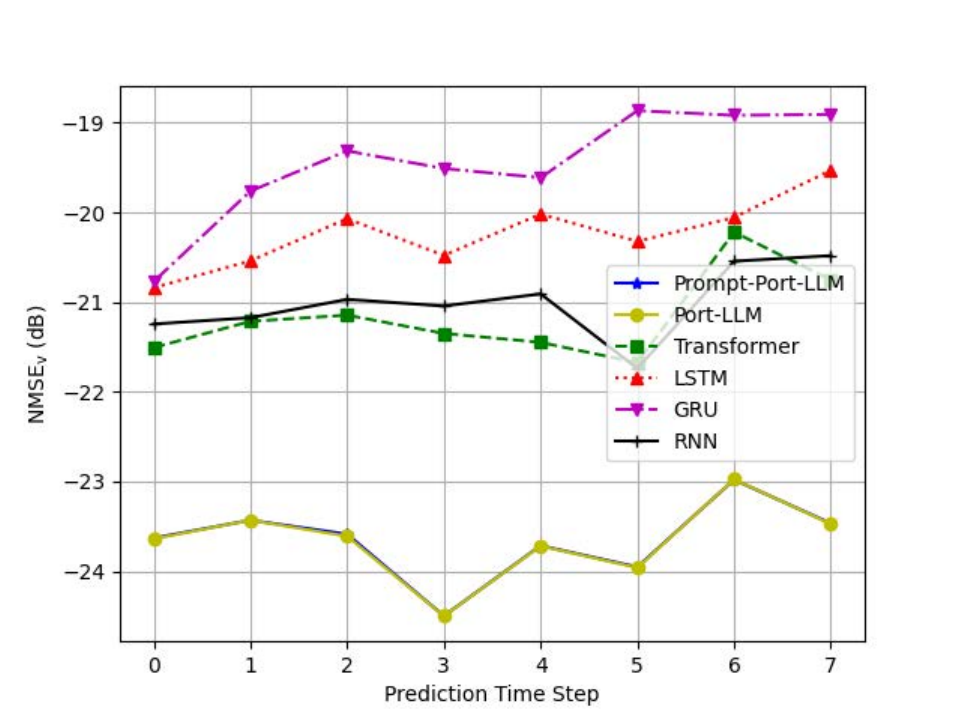}
    \end{minipage}
}
\subfigure[]
{
 	\begin{minipage}[b]{.31\linewidth}
        \centering
        \includegraphics[scale=0.40]{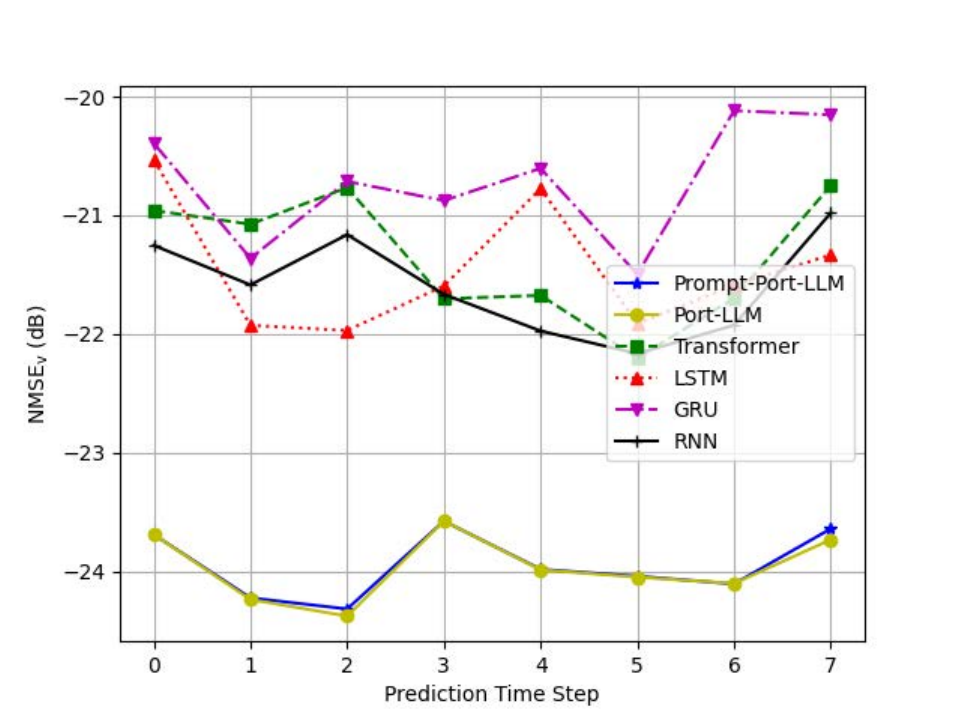}
    \end{minipage}
}
\caption{When the number of antennas on the BS side is $32\times 8$, the performance of different models under various velocities. (a) The test velocity is 90 km/h; (b) The test velocity is 120 km/h; (c) The test velocity is 150 km/h.}
\label{fig:different_models_32_8}
\end{figure*}
\begin{figure}[t!]
    \centering
    \includegraphics[scale=0.65]{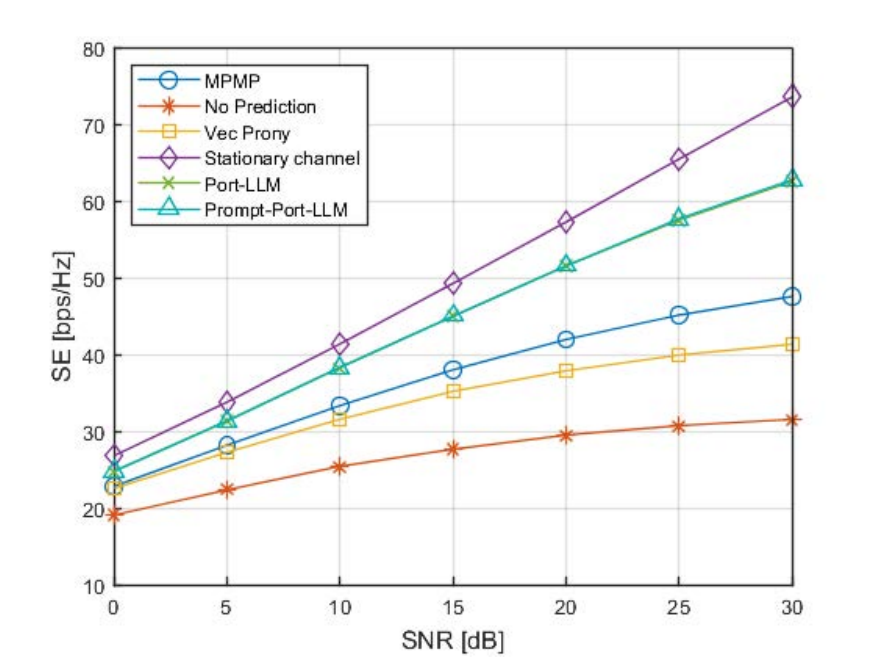}
    \caption{The SE versus SNR, the BS has $2\times 8$ antennas, the velocity of UE is 90 km/h.}
    \label{fig:se_2_8_v90}
\end{figure}
\begin{figure}[t!]
    \centering
    \includegraphics[scale=0.65]{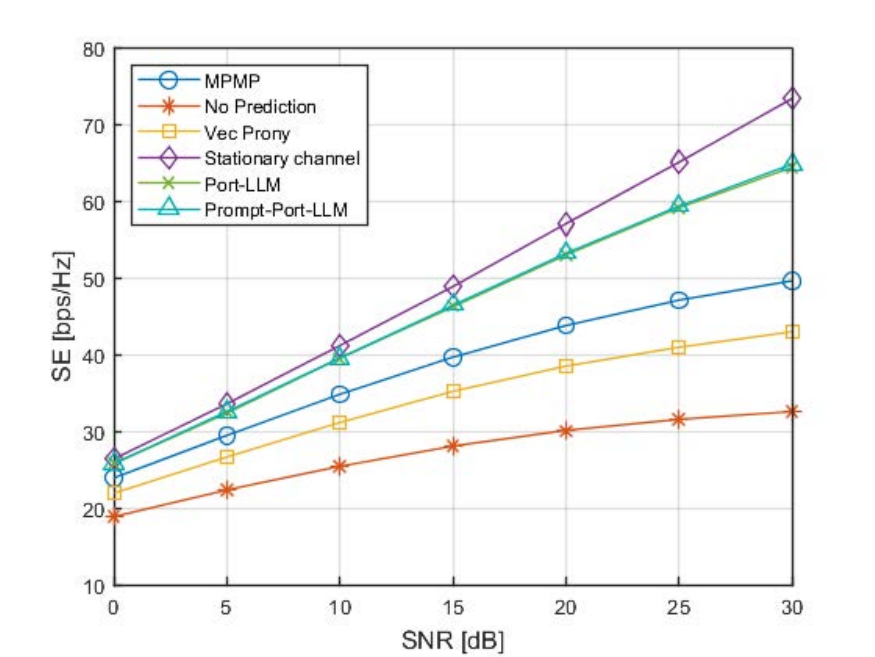}
    \caption{The SE versus SNR, the BS has $2\times 8$ antennas, the velocity of UE is 120 km/h.}
    \label{fig:se_2_8_v120}
\end{figure}
Fig. \ref{fig:different_models_2_8}, Fig. \ref{fig:different_models_8_8}, and Fig. \ref{fig:different_models_32_8} provide a comparative analysis of the efficacy of our model in predicting FA moving ports relative to four other neural network-based models. This evaluation is conducted across three distinct BS antenna configurations and three varying UE mobility speeds. The horizontal axis of the figures denotes 8 consecutive prediction moments, while the vertical axis represents the NMSE between the predicted channels of moving ports and the reference channels. In each scenario, UE data is collected from 10 different orientations, with each UE contributing data over 50 consecutive moments, specifically sampling at the 7-th moment within each time period. The NMSE values predicted by the model are subsequently averaged across all data. The analysis presented in these figures indicates that our proposed Port-LLM model and Prompt-Port-LLM model exhibit superior predictive performance, surpassing that of the RNN-based, LSTM-based, GRU-based and Transformer-based model. This deficiency can be attributed to the limited modeling capabilities of the RNN-based, LSTM-based, GRU-based and Transformer-based architectures when addressing complex sequential challenges. Figs. 8-10 also show that our proposed model trained in the SISO scenario directly applies to the MISO scenario also has good prediction performance. It is important to highlight that, as illustrated in Fig. 6 and Figs. 8-10, although the accuracy of the channel table predicted by our proposed Prompt-Port-LLM model is slightly higher than that of our proposed Port-LLM model (about 1$\%$ higher), the NMSE between the channels corresponding to the predicted moving ports and the reference channels for both models is almost the same, approximately -24 dB. As shown in Figs. 8-10, the corresponding curves for both our proposed models are basically overlapping. This phenomenon can be attributed to the fact that there exists a certain ``tolerance range" in selecting the moving port from the channel table that best matches the reference channel for a fixed density of moving ports (i.e., channel table size) of the FA. Tolerance range means that we only need to ensure that, within the channel table predicted by our model, the channel corresponding to the predicted moving port is the closest to the reference channel. It is not necessary to account for the precise proximity between these two channels. Therefore, even if the prediction accuracy of the Prompt-Port-LLM model is slightly better, the moving ports predicted by both models may still be highly consistent. This further demonstrates that the prediction accuracy of our proposed Port-LLM model, based on LoRA fine-tuning, is already very high for our model task.\par

\begin{figure}[t]
    \centering
    \includegraphics[scale=0.65]{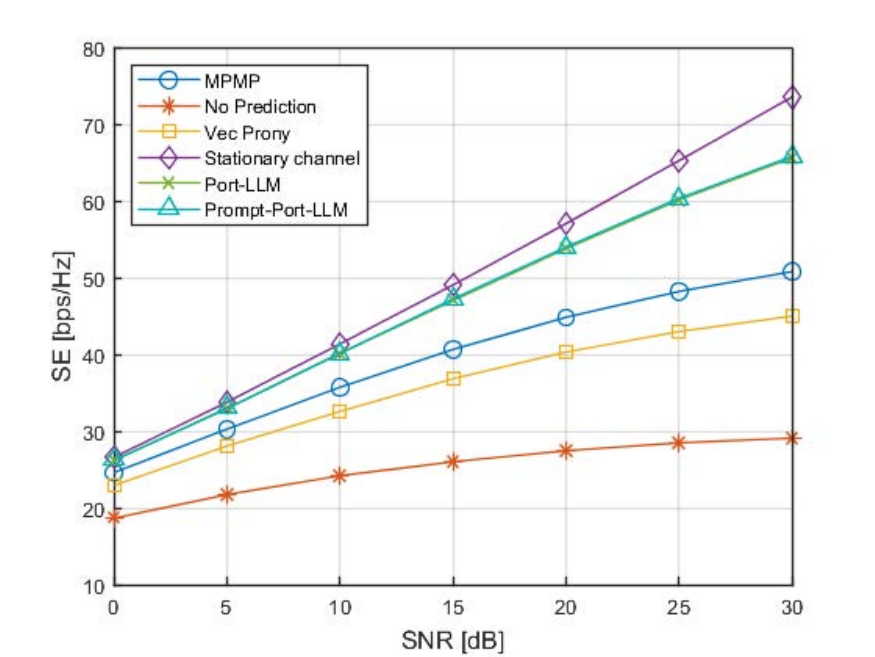}
    \caption{The SE versus SNR, the BS has $2\times 8$ antennas, the velocity of UE is 150 km/h.}
    \label{fig:se_2_8_v150}
\end{figure}
Similarly, with a BS antenna configuration of $2\times 8$ and 10 UEs considered, we conduct a comparative analysis of the SE achieved by our proposed Port-LLM model and Prompt-Port-LLM model against the SE derived from both the Vec Prony algorithm and the MPMP algorithm. Additionally, we also evaluat the SE under the idealized scenario (``Stationary channel") and the absence of channel prediction (``No Prediction"). As illustrated in Fig. \ref{fig:se_2_8_v90}, Fig. \ref{fig:se_2_8_v120}, and Fig. \ref{fig:se_2_8_v150}, optimal performance is observed under ``Stationary channel" condition. Conversely, performance is significantly diminished in the ``No Prediction" condition. Furthermore, the Port-LLM model and the Prompt-Port-LLM model we proposed achieve better SE than those obtained using the MPMP algorithm and the Vec Prony algorithm. It indicates that, compared to the MPMP algorithm and the Vec Prony algorithm, both of our proposed models are more effective in accurately capturing UE mobility in the context of UE movement, resulting in more accurate predictions of the moving ports of the FA at future moments. Additionally, the consideration of a two-dimensional (2D) region for FA movement in our proposed models, as opposed to the one-dimensional (1D) framework utilized by the MPMP algorithm, further enhances the performance of our models.\par

\begin{table}[t]
  \centering
  \caption{The $\text{Accuracy}$, $\text{NMSE}_{\text{v}}$, network parameters, and interference time of different models.}
  \label{tab:training_parameter}
  \renewcommand{\arraystretch}{1.5}
  \begin{tabularx}{\linewidth}{>{\centering\arraybackslash}X >{\centering\arraybackslash}X >{\centering\arraybackslash}X >
  {\centering\arraybackslash}X >
  {\centering\arraybackslash}X} 
    \toprule
    Model & $\text{Accuracy}$ (\%) & $\text{NMSE}_{\text{v}}$ (dB) & Network parameters ($1\times 10^{6}$) & Interference time (ms) \\
    \midrule
    LSTM-based & 91.16 & -19.64 & 242.46/242.46 & $-$\\
    RNN-based & 93.15 & -20.96 & 235.37/235.37 & $-$\\
    GRU-based & 92.94 & -20.60 & 240.10/240.10  & $-$\\
    Transformer-based & 93.40 & -21.01 & 238.12/238.12 & $-$\\
    Port-LLM & 98.18 & -23.89 & 232.70/313.80 & 4.72\\
    % Port-LLM (Fine Tuning) & 98.45 & 313.73/313.73 & 4.81\\
    Prompt-Port-LLM & 99.45 & -23.91 & 233.00/314.92 & 7.52\\
    \bottomrule
  \end{tabularx}
\end{table}

In order to assess the deployment feasibility of our proposed models in real applications, we investigate the number of training parameters for our proposed Port-LLM model, Prompt-Port-LLM model, and other comparative neural network-based models. In addition, we also investigate the inference time required for our proposed models to perform one prediction. All of our neural network-based comparison experiments are conducted on a server equipped with 1 NVIDIA GeForce RTX 3090 GPU, 1 Intel Core i9-10920X CPU and 94 GB of RAM. As illustrated in Table III, the predictive performance of our models surpasses that of the other neural network-based models. Although the model parameters of the proposed Port-LLM and Prompt-Port-LLM models exceed those of other neural network-based models, it is important to note that a portion of these parameters are frozen, resulting in the number of parameters requiring retraining that is comparable to that of other neural network models. Furthermore, both the Port-LLM and Prompt-Port-LLM models achieve milliseconds of inference speed. Additionally, the reasoning speed of the Port-LLM model is faster than that of the Prompt-Port-LLM model. This discrepancy can be attributed to the fact that the Prompt-Port-LLM model necessitates additional time to acquire dynamic prompt prior to each prediction, whereas the Port-LLM model, which utilizes LoRA fine-tuning, does not have this requirement.\par

\section{Conclusions}\label{conclusion}
In this paper, the FA is employed to mitigate mobility-induced challenges in communication systems. In particular, leveraging the powerful modeling capabilities of LLMs, we propose a Port-LLM model based on LoRA fine-tuning. Building on this, we also propose a Prompt-Port-LLM model based on prompt fine-tuning to further utilize the exceptional NLP capabilities of LLMs. By repositioning the FA to the port predicted by our proposed models, it becomes feasible to maintain an approximately invariant channel state information as the UE moves. To effectively address the discrepancies between the wireless communication data format and the input format of the pre-trained LLM, we specially design the data processing module, input embedding module and output projection module. Moreover, in our proposed Prompt-Port-LLM model, we design specialized dynamic prompts and a dedicated prompt encoder module, thereby further enhancing the model's prediction accuracy. Simulation results show that both of our proposed models exhibit significant performance improvement compared to the traditional technical solutions, especially in medium and high-speed scenarios. In addition, our proposed models show good robustness under different BS-side antenna configurations and different UE movement speed conditions. In terms of computational efficiency, the inference speed of both models reaches the millisecond level.\par

%%%%%%%%%%%%%%%%%%%%%%%%%%%%%%%%%%%%% %\renewcommand{\baselinestretch}{1.3525}
\bibliographystyle{IEEEtran}
\bibliography{ref}

\end{document}